\DocumentMetadata{}
\documentclass[manuscript,screen]{acmart}

\AtBeginDocument{%
  \providecommand\BibTeX{{%
    \normalfont B\kern-0.5em{\scshape i\kern-0.25em b}\kern-0.8em\TeX}}}

\setcopyright{acmcopyright}
\copyrightyear{2025}
\acmYear{2025}
\acmDOI{XXXXXXX}

\acmConference[Woodstock '18]{Woodstock '18: ACM Symposium on Neural
  Gaze Detection}{June 03--05, 2018}{Woodstock, NY}
\acmBooktitle{Woodstock '18: ACM Symposium on Neural Gaze Detection,
  June 03--05, 2018, Woodstock, NY}
\acmPrice{15.00}
\acmISBN{978-1-4503-XXXX-X/18/06}

\usepackage{xspace}

\usepackage{xcolor}
\usepackage{ifthen}
\usepackage[resetlabels,labeled]{multibib}
\usepackage{multirow}
\usepackage{xparse}
\usepackage{float}
\usepackage{pgfplots}
\usepackage{tikz}
\usepackage{pifont}% http://ctan.org/pkg/pifont
\usepackage{fontawesome}
\newcommand{\cmark}{\ding{51}}%
\newcommand{\xmark}{\ding{55}}%

\usepackage{soul} %Strikethrough text

\usepackage{hhline,graphicx,array}
\usepackage{tabularx,colortbl}
\definecolor{my_color}{rgb}{220,220,220}
\definecolor{lightlightgrey}{RGB}{233, 236, 239}

\newcites{P}{List of Selected Studies}
\let\oriCiteP\citeP
\RenewDocumentCommand{\citeP}{O{} O{} m}{%
  \renewcommand{\citenumfont}[1]{P##1}%  
  \oriCiteP[#1][#2]{#3}%
  \renewcommand{\citenumfont}[1]{##1}%
}

\newcommand{\ie}{\emph{i.e.,}\xspace}
\newcommand{\eg}{\emph{e.g.,}\xspace}
\newcommand{\etc}{etc.\xspace}
\newcommand{\etal}{\emph{et~al.}\xspace}
\newcommand{\secref}[1]{Section~\ref{#1}\xspace}

\newcommand{\figref}[1]{Fig.~\ref{#1}\xspace}

\newcommand{\tabref}[1]{Table~\ref{#1}\xspace}

\newboolean{showcomments}
\setboolean{showcomments}{true} % toggle to show or hide comments
\ifthenelse{\boolean{showcomments}}
  {
   
  }
  {\newcommand{\nbb}[2]{}

  }
  
\newcommand{\selected}{119\xspace} %updated
\newcommand{\queried}{9,845\xspace} %updated

\newcommand{\numberOfTasks}{34\xspace} %updated

\begin{document}

\title{Automating Code Review: A Systematic Literature Review}

\author{Rosalia Tufano}
\affiliation{%
  \institution{SEART @ Software Institute - Universit\`{a} della Svizzera italiana}
  \country{Switzerland}
}

\author{Gabriele Bavota}
\affiliation{%
  \institution{SEART @ Software Institute - Universit\`{a} della Svizzera italiana}
  \country{Switzerland}
}

\renewcommand{\shortauthors}{Tufano and Bavota}

\begin{abstract}
% No more than 200 words
Code Review consists in assessing the code written by teammates with the goal of increasing code quality. Empirical studies documented the benefits brought by such a practice that, however, has its cost to pay in terms of developers' time. For this reason, researchers have proposed techniques and tools to automate code review tasks such as the reviewers selection (\ie identifying suitable reviewers for a given code change) or the actual review of a given change (\ie recommending improvements to the contributor as a human reviewer would do). Given the substantial amount of papers recently published on the topic, it may be challenging for researchers and practitioners to get a complete overview of the state-of-the-art.

We present a systematic literature review (SLR) featuring \selected papers concerning the automation of code review tasks. We provide: (i) a categorization of the code review tasks automated in the literature; (ii) an overview of the under-the-hood techniques used for the automation, including the datasets used for training data-driven techniques; (iii) publicly available techniques and datasets used for their evaluation, with a description of the evaluation metrics usually adopted for each task.

The SLR is concluded by a discussion of the current limitations of the state-of-the-art, with insights for future research directions.

\end{abstract}

%%
%% The code below is generated by the tool at http://dl.acm.org/ccs.cfm.
%% Please copy and paste the code instead of the example below.
%%



\maketitle

%!TEX root = main.tex

%%%%%%%%%%%%%%%%%%%%%%
%%%%%%%%%%%%%%%%%%%%%%
\section{Introduction}  \label{sec:intro}
%%%%%%%%%%%%%%%%%%%%%%
%%%%%%%%%%%%%%%%%%%%%%

The idea of inspecting peers' code looking for bugs and suboptimal implementation choices dates back to the 70s and in particular to the seminal work by Fagan titled ``\emph{Design and code inspections to reduce errors in program development}'' \cite{fagan:ibm1976}. The formal code inspections envisioned at that time slowly evolved into what is know as \emph{modern code review} (MCR) \cite{bacchelli:icse2013}, being tool-based and more informal.

One of the objectives of MCR is to reduce the inherent cost associated with code review. Indeed, while there is ample evidence about the benefits of code review \cite{bacchelli:icse2013,mcintosh:msr2014,morales:saner2015,bavota:icsme2015,sadowski:icse2018}, they do not come for free, and may result in developers spending many hours per week reviewing code \cite{bosu:esem2013}.

For this reason, researchers proposed techniques and tools to automate specific code review tasks. For example, several studies focus the attention on the task of \emph{recommending reviewers} %\textcolor{red}{\reviewerRecommendation task} 
\cite{balachandran:icse2013, jiang:jcst2015, thongtanunam:saner2015, xia:icsme2015, ouni:icsme2016, ying:csi2016, yu:ist2016, zanjan:tse2016, xia:sm2017,  jiang:ist2017, fejzer:jiis2018, asthana:esecfse2019, liao:globecom2019, sulun:icpmdase2019, jiang:jss2019, mirsaeedi:icse2020, al:icpmda2020, strand:icse-sep2020, chouchen:asc2021, tecimer:ease2021, pandya:esecfse2022, li:ease2023, aryendu:ase2023, zhang:icse2023, rahman:icse2016, chueshev:icsme2020, rebai:ase2020, ye:ieee2019, zhao:cascon:2022, qiao:saner2024, hajari:tse2024, rahman:icsme2023, rong:icse2022, sulun:ist2021, kong:saner2022, ahasanuzzaman:emse2024}, namely the automatic selection of proper reviewers for a given code change. Other researchers target instead the task of classifying reviewers' comments %\commentsClassification task 
\cite{li:seke2017, fregnan:emse2022}, having the goal of automatically classifying comments posted by reviewers based on the ``type of feedback'' they provide to the contributor (\eg feedback about the code \emph{style}, \emph{functionality}, \etc). With the recent adoption of deep learning (DL) in software engineering, generative tasks have also been subject of automation. For example, DL models have been trained with the goal of generating natural language comments asking to the contributor code changes as a human reviewer would do (\ie simulating a reviewer commenting on the submitted code) \cite{li.l:esecfse2022, li:fse2022, hong:esecfse2022, tufano:icse2022, vijayvergiya:aiware2024, lu:ase2024, yu:tosem2024, lin:msr2024, sghaier:fse2024, lu:issre2023, nashaat:tse2024}. 

Given the numerous code review tasks in which automation attempts have been made and the large number of studies targeting this topic, it is important to synthesize the current state-of-the-art to provide researchers and practitioners with an updated entry point on code review automation. 

We present a systematic review of the literature presenting techniques and tools for the automation of code review tasks. Previous secondary studies on the topic \cite{hannebauer:ase2016,coelho:iwor2019} only focused on specific tasks (\ie \emph{recommending reviewers} task and refactoring-aware solutions) or do not have a specific focus on code review automation \cite{davila:jss2021}. As we show, there are \numberOfTasks tasks for which researchers proposed automated solutions in \selected articles. As a comparison, the most extensive literature review at date also featuring code review automation techniques only includes 53 of these articles \cite{davila:jss2021}. This makes our SLR by far the most comprehensive at date on the topic of code review automation. The SLR we present is the result of filtering out \selected relevant studies out of 11,165 resulting from querying popular digital libraries. Our contributions are: (i) A categorization of the \numberOfTasks code review tasks for which researchers proposed automated solutions; (ii) An overview of the techniques used in the literature to automate code review (\eg exploiting machine learning, DL, information retrieval, \etc) with a focus on the training strategies used for data-driven techniques; (iii) A collection of the publicly available techniques (\ie the tool or the code implementing the technique is publicly available) and evaluation datasets clustered by ``type of automated tasks'' (\eg we list all publicly available tools/techniques and evaluation datasets for the task of \emph{recommending reviewers}); (iv) A description of the evaluation frameworks adopted in the literature to assess the performance of techniques proposed for the different tasks, with a focus on the adopted metrics, targeted language, and deployment in industry of the automated solution; (v) Informed by the finding of our SLR, we list directions for future work in the field of code review automation.

%%%%%%%%%%%%%%%%%%%%%%%%%%%%%%%%%%%
\subsection{Structure of the Paper}
%%%%%%%%%%%%%%%%%%%%%%%%%%%%%%%%%%%

\secref{sec:related} reports the related literature, presenting surveys, SLR and mapping studies dealing with modern code review. \secref{sec:method} presents the methodology we adopted to conduct the SLR. \secref{sec:results} discusses the achieved results, answering our research questions. \secref{sec:threats} reports the threats that could affect the validity of our findings. Finally, \secref{sec:conclusion} concludes the paper.

%!TEX root = main.tex

%%%%%%%%%%%%%%%%%%%%%%
%%%%%%%%%%%%%%%%%%%%%%
\section{Related Work}  \label{sec:related}
%%%%%%%%%%%%%%%%%%%%%%
%%%%%%%%%%%%%%%%%%%%%%

\begin{table*}[ht]
\centering
    \caption{Surveys, SLRs and mapping studies dealing with modern code review}
    \label{tab:related}
    {\footnotesize
    \begin{tabular}{@{}lp{7cm}rrr@{}}
    \toprule
    \textbf{Reference}    & \textbf{Main Goal} & \textbf{Year} & \textbf{\#Papers} & \textbf{\#Papers Automation}\\ \midrule
    Hannebauer \etal \cite{hannebauer:ase2016} & Comparing eight techniques to recommend code reviewers & 2016 & 8 & 8\\\midrule
    
    Badampudi \etal \cite{badampudi:ease2019} & Documenting the research questions addressed in code review literature & 2019 & 177 & 39\\\midrule

	Coelho \etal \cite{coelho:iwor2019} & Mapping refactoring-aware solutions to support modern code review & 2019 & 13 & 9\\\midrule
    
    Fronza \etal \cite{fronza:2020} & Documenting the research questions addressed in code review literature & 2020 & 75 & 0\\\midrule
    
    Wang \etal \cite{wang:jss2021} & Mapping the type of contributions (\eg empirical study, automation) in code review papers, study their replicability, document the type of data collected in such studies (\eg the experience of reviewers, the workload, \etc) & 2021 &  112 & 37\\\midrule
    
    Davila \etal \cite{davila:jss2021} & Mapping the type of contributions (\ie foundational, proposals, evaluations) in code review papers & 2021 & 139 & 53\\\midrule
    
    \textbf{Our work} & \textbf{Documenting the code review tasks automated in the literature, the adopted techniques and evaluation datasets/frameworks} & \textbf{2025} & \textbf{\selected} & \textbf{\selected}  \\ \bottomrule
    \end{tabular}
    }
\end{table*}

\tabref{tab:related} lists the previous secondary studies on modern code review in chronological order. For each work we also include (i) the overall number of papers part of the study (\ie column ``\#Papers'') and (ii) the papers related to the automation of code review tasks that are featured in the study. 

The focus of the works by Badampudi \etal \cite{badampudi:ease2019}, Wang \etal \cite{wang:jss2021}, and Fronza \etal \cite{fronza:2020} is different as compared to our SLR.  Badampudi \etal \cite{badampudi:ease2019} aim at classifying the literature on modern code review based on the investigated research questions. As a result of this analysis they report that 39 of the surveyed papers present tool support for code review. However, no additional analyses are performed on these works. A similar study has also been presented by Wang \etal \cite{wang:jss2021}. Also in this case the authors focus on classifying the type of contribution, reporting 37 papers out of the 112 considered as related to code review automation. Again, there are not specific research questions in the SLR about code review automation.
Fronza \etal \cite{fronza:2020}, instead, explicitly focus on empirical studies rather than papers presenting techniques and tools for code review automation. 

Hannebauer \etal \cite{hannebauer:ase2016} and Coelho \etal \cite{coelho:iwor2019} present secondary studies focusing on the automation of specific code review tasks. The former compares eight techniques for the \emph{recommending reviewers} task, while the latter looks at 13 refactoring-aware solutions proposed in the literature. Our SLR has a wider target, looking at works automating any code review task.

Finally, Davila \etal \cite{davila:jss2021} presented another SLR mapping the type of contribution of the code review papers. Their SLR features 53 papers presenting tools and techniques for the automation of code review. As compared to the previously discussed SLRs, Davila \etal provide a detailed description of these works, including the type of task they support. However, differently from our SLR, the main focus is not on code review automation and, due to the time period in which papers have been collected (\ie up to 2019), none of the recent techniques built on top of DL models is documented (and, consequently, none of the tasks that have been automated for the first time thanks to DL models). Our SLR more than doubles the paper on code review automation present in the work by Davila \etal \cite{davila:jss2021} (from 53 to \selected).
%!TEX root = main.tex

%%%%%%%%%%%%%%%%%%%%%%%%%
%%%%%%%%%%%%%%%%%%%%%%%%%
\section{Research Method} \label{sec:method}
%%%%%%%%%%%%%%%%%%%%%%%%%
%%%%%%%%%%%%%%%%%%%%%%%%%

We describe our research method following the guidelines by Kitchenham and Charters \cite{kitchenham:2007} for SLR.

%%%%%%%%%%%%%%%%%%%%%%%%%%%%%%%
\subsection{Research Questions} \label{sec:rqs}
%%%%%%%%%%%%%%%%%%%%%%%%%%%%%%%
Our SLR aims at informing researchers and practitioners about the state of the art in automating code review and it is thus steered by the following research questions (RQs):

\begin{itemize}
\item \textbf{RQ$_1$:} \emph{What are the code review tasks for which researchers proposed automated solutions?} We aim at categorizing the code review tasks automated in the literature to support (i) researchers, in getting a complete overview of tackled research directions in the field, thus possibly identifying areas in needed of further research; and (ii) practitioners, in discovering automated solutions which may be employed in their daily workflow.
\end{itemize}

Once identified the list of automated tasks, the following RQs are answered for each task (\ie by discussing the findings by task):

\begin{itemize}
\item \textbf{RQ$_2$:} \emph{What are the under-the-hood solutions behind the techniques and tools proposed for code review automation?} This RQ sheds light on the functioning of the proposed automated solutions. In particular, we present: (i) a high-level classification of the adopted technical solution --- \eg DL-based, ML-based (Machine Learning), IR-based (Information Retrieval), \etc; (ii) a description of the  training strategies adopted in data-driven solutions; and (iii) information about the programming language target of the automation (\eg does the technique only support Java or it is language-independent?).

\item \textbf{RQ$_3$:} \emph{How are techniques for the automation of code related tasks empirically evaluated?} We focus on the adopted evaluation metrics and on additional qualitative/industrial studies present in the papers. RQ$_3$ can help researches in getting a quick understanding of the possible evaluation framework to adopt for their techniques. 

\item \textbf{RQ$_4$:} \emph{Which techniques and datasets are publicly available?} While RQ$_1$ identifies the  automated tasks and, for each of them, lists the solutions proposed in the literature, not all these techniques are publicly available (\ie their implementation has been released by the authors). A similar observation can be made for the used datasets. The output of RQ$_4$ is the list of techniques and datasets that, as of the day of writing (January 2025), are publicly available. Such an outcome can be useful to (i) researchers, to easily identify baselines for comparisons and/or datasets that can be used for building or evaluating automated solutions; and (ii) practitioners, to easily spot ``ready-to-use'' solutions they can consider for adoption.

\item \textbf{RQ$_5$:} \emph{What are the concerns raised or the limitations observed by researchers when experimenting the automated solutions?} We inspect the \selected papers to identify limitations and concerns researchers discuss about the proposed techniques, with the goal of outlining possible future research directions in the field.

\end{itemize}

%%%%%%%%%%%%%%%%%%%%%%%%%%%%%%%%%%%%%%%%%%
\subsection{Relevant Study Identification}
%%%%%%%%%%%%%%%%%%%%%%%%%%%%%%%%%%%%%%%%%%

\begin{figure*}[]
	\centering
	\includegraphics[width=0.8\linewidth]{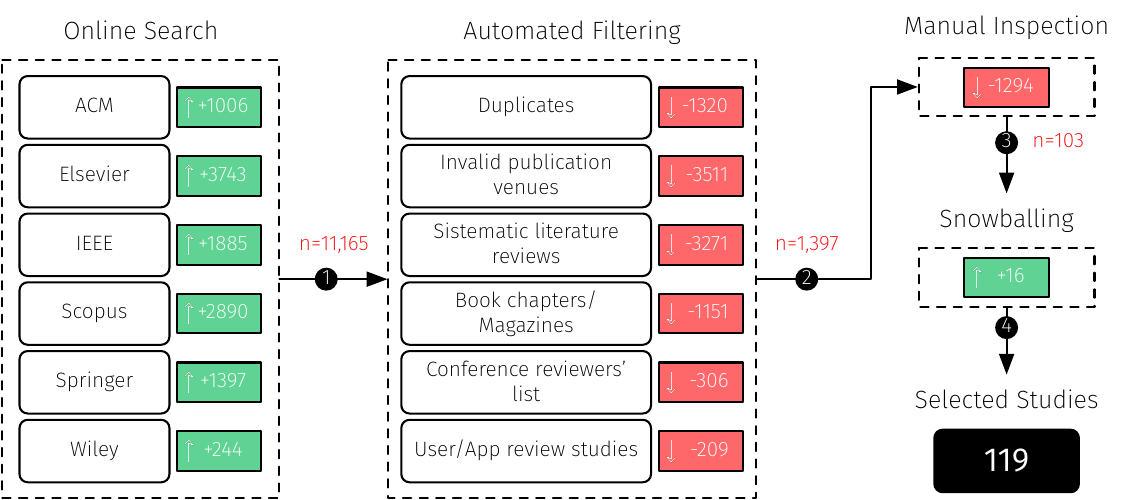}
	\caption{Study selection process}
	\label{fig:process}
\end{figure*}

\figref{fig:process} depicts the process adopted to identify the relevant primary studies. Such a process is detailed in the following. 

\subsubsection{Search Strategy} We queried six digital libraries to search for primary studies: ACM Digital Library \cite{ACM}, Elsevier ScienceDirect \cite{Elsevier}, IEEE Xplore Digital Library \cite{IEEE}, Scopus \cite{Scopus}, Springer Link Online Library \cite{Springer}, and Wiley Online Library \cite{Wiley}. We did not query Google Scholar due to the limitations documented by Halevi \etal \cite{halevi:ji2017} (\eg lack of quality control, missing support for data download). 

To define the query needed to identify works related to the automation of code review tasks, a trial-and-error procedure has been performed by the two authors. It became soon clear that searching in the paper titles for keywords such as ``\emph{automating}'', ``\emph{recommending}'', \etc was not an option, even considering all their possible variations (\eg \emph{automating}, \emph{automated}, \emph{automate}). Indeed, this would have led to the lost of several relevant studies (\eg ``\emph{Intelligent Code Review Assignment for Large Scale Open Source Software Stacks}'' \cite{aryendu:ase2023}, ``\emph{A Multi-Step Learning Approach to Assist Code Review}'' \cite{sghaier:saner2023}). For this reason, we opted for a more conservative query which targets the identification of all code review-related studies, even those do not presenting automated solutions:

\begin{flushleft}
\small
\textbf{Title} CONTAINS\\
\hspace{1.1cm}``\emph{revi*}'' OR (``\emph{cod*}'' AND ``\emph{edit*}'') AND\\
\textbf{Publication venue} CONTAINS\\
\hspace{1.1cm}(``\emph{software}'' OR ``\emph{program}'' OR ``\emph{code}'')
\end{flushleft}

The query searches for the term ``\emph{revi*}'' (\eg review, reviewing, revision) or both the terms ``\emph{cod*}'' (\eg code, coding) and ``\emph{edit*}'' in the article title. The latter have been included to match works related to the recent trend of automating code editing needed to address a reviewer's comment (see \eg \cite{tufano:icse2021, tufano:icse2022, huq:ist2022, li:fse2022, sghaier:fse2024, lu:issre2023, nashaat:tse2024, zhang:ase2022, pornprasit:ist2024, lu:apsec2023, froemmgen:icse2024}). While only searching in the title might be restrictive, we want to identify automated solutions which have been explicitly proposed for code review (\eg we are not interested in articles presenting generic static analysis tools that might be applied in code review to spot quality issues). Also, we only searched for articles published in venues containing at least one of three keywords: ``\emph{software}'', ``\emph{program}'', and ``\emph{code}''. Such a filter is based on the authors' knowledge of software engineering publication venues. We acknowledge that there might be relevant articles published in related fields (\eg artificial intelligence) that our query would exclude. However, as explained later, we adopt a snowballing process to partially address this issue. 

Among the queried search engines Elsevier, Scopus, Springer, and Wiley allow to specify a \emph{discipline} of interest, which is useful to minimize the retrieved false positive instances. For these libraries, we selected ``Computer Science'' as discipline. Springer also allows to specify sub-disciplines, for which we selected ``Software Engineering/Programming''. The link with the query used for each digital library is publicly available in our replication package \cite{replication}. The query has been run on 20 December 2024 on all digital libraries. 

\begin{table}[ht]
\centering
    \caption{Articles returned by the queried digital libraries}
    \label{tab:search}
    {\footnotesize
    \begin{tabular}{@{}lr@{}}
    \toprule
    \textbf{Source}                      & \textbf{Returned Articles} \\ \midrule
    ACM Digital Library               &              1,006   \\
    Elsevier ScienceDirect          &              3,743 \\
    IEEE Xplore Digital Library   &              1,885\\
    Scopus                                  &              2,890\\
    Springer Link Online Library &               1,397\\
    Wiley Online Library              &               244\\\midrule
    Total (including duplicates)    &               11,165       \\ \midrule
    \textbf{Total (excluding duplicates)}   & \textbf{\queried}                \\ \bottomrule
    \end{tabular}
    }
\end{table}

\tabref{tab:search} reports the articles returned by each digital library. Once removed duplicates (\ie the same article has been returned by multiple libraries), we collected \queried candidate primary studies which have been manually inspected as described in the following. 

\subsubsection{Study Selection}
\label{sec:study_select}
Given the high number of articles returned by the formulated query, we started with an automated check aimed at excluding clear false positives. First, despite the filter on venues we set in the digital libraries, we noticed that some of the returned results concerned invalid publication venues (\ie venues not featuring in their name any of the three keywords ``\emph{software}'', ``\emph{program}'', and ``\emph{code}''). Thus, we implemented a simple script excluding those cases (-3,511). 

Other three filters were implemented. First, given our query, and in particular the retrieval of articles containing ``\emph{revi*}'' in their title, we retrieved several SLRs. Among those, we were only interested in the ones focusing on code review, since they represent an important source of references for the snowballing phase. Thus, we automatically removed all articles containing in the title, besides ``\emph{review}'', the term ``\emph{systematic}'' and do not containing the term ``\emph{code}'' (-3,271). Second, we excluded articles published as book chapters or in magazines, since those are usually not full research articles (-1,151). Finally, we also excluded ``reviewers lists'' (-306) and works related to user/app reviews (-209). At the end of this process, 1,397 candidate primary studies were left. 

\begin{table*}[ht]
	\centering
    \caption{Inclusion and exclusion criteria}
    \label{tab:criteria}
    {\footnotesize
    \begin{tabular}{@{}lp{14cm}@{}}
    \toprule
    & {\bf Inclusion Criteria}\\
    \midrule
    
    IC1  & The article must be peer-reviewed, published at conferences, workshops, or journals. In the snowballing phase later described, we ignore all referenced preprints (\eg those published on arXiv.org).
    \\
    
    IC2  & The PDF of the article must be available online. We searched for it on the online libraries featuring and, if needed, on Google. 
    \\
    
    IC3 & The article must present technique(s) to automate a code review task. It is not enough to present a generic technique that, accordingly to the reader, \emph{might} be useful in the context of code review: The authors must explicitly state that the technique has been thought to support code review. \\
    
    \midrule
    & {\bf Exclusion Criteria}\\
    \midrule
        EC1  & The article is not written in English.
    \\
    EC2 & The article has been published in a conference/workshop and later on extended to a journal. We only keep the journal article to avoid redundancy. 
    \\
    EC3 & The article is not a full research publication (\eg doctoral symposium articles, posters, ERA track). We exclude all articles having less than six pages with the goal of removing articles that may not have been subject to the same peer-review process typical of full research articles.
    \\ 
    EC4 & The article replicates a previously published technique for code review automation which has been already included in the SLR.
    \\ 
    EC5 & The article is a secondary study. In this case, we keep it only as a source of references for the snowballing phase.
    \\ 
    EC6 & The article has not been published in an international venue, but in a national one (\eg \emph{Brazilian Symposium on Programming Languages}).
    \\ \bottomrule
    \end{tabular}
    }
\end{table*}

This set has then been manually inspected by both authors. Inclusion and exclusion criteria are listed in \tabref{tab:criteria}. This part of the manual analysis was mainly focused on the inspection of the title and abstract of the article. Authors agreed to be conservative and include the article in case of doubts, given the planned subsequent reading of the whole article as described in the following.  Conflicts (\ie cases in which one author considered the article as relevant and one not) arisen in 25 cases (1.8\%) and have been solved through an open discussion. This filtering process left 175 candidate studies which have been equally split among the two authors. Each author downloaded the corresponding article and re-inspected it keeping the inclusion and exclusion criteria in mind (\tabref{tab:criteria}) and then either confirming the article as relevant for the SLR or discarding it. All those discarded have been double-checked by the other author to ensure no relevant studies were mistakenly excluded. 

This further check confirmed 103 articles as relevant primary studies. Those, together with 19 articles tagged as ``relevant secondary study'', have been subject of a backward snowballing process.

\textbf{Backward Snowballing.} The 122 articles were equally split among the authors, with each of them in charge of reading the reference list and identify possible relevant papers. At this step, we retrieved also relevant papers published in venues not containing any of the three keywords ``\emph{software}'', ``\emph{program}'', and ``\emph{code}'' (\eg papers published in the \emph{Conference on Artificial Intelligence --- AAAI}). Also in this phase, in case of doubts, the authors agreed to included a referenced article for a further check by the other author. The snowballing resulted in 16 additional primary studies, that summed up to the 103 previously collected leads to the final set of \selected primary studies featured in our SLR.

\begin{table}[ht]
	\centering
    \caption{Data extraction questionnaire}
    \label{tab:data}
    {\footnotesize
    \begin{tabular}{@{}lp{14cm}l@{}}
    \toprule
    {\bf No.} & {\bf Question} & {\bf Focus}\\
    \midrule
    Q1 & Which code review task has been automated? & RQ$_1$\\\midrule
    Q2 & Does the employed technique rely on machine/deep learning? & RQ$_2$\\
    Q3 & If yes to Q2, which specific algorithms are used? & RQ$_2$\\ 
    Q4 & If no to Q2, summarize the approach functioning. & RQ$_2$\\
    Q5 & Which dataset has been exploited to build the technique?\footnote{We use ``build'' instead of ``train'' since the technique might not exploit any sort of learning but being, for example, and IR-based approach employing a dataset as the source from which retrieving relevant pieces of information.} Collect information related to (i) the subject programming language(s) and (ii) the type of information featured in the dataset (\ie what is an ``instance'' in the dataset?). & RQ$_2$\\\midrule
    Q6 & Which evaluation metrics have been employed? & RQ$_3$\\
    Q7 & Did the authors perform any sort of qualitative analysis? & RQ$_3$\\
    Q8 & Was the approach deployed in an industrial setting? & RQ$_3$\\\midrule
    Q9 & Is a link to a replication package available? Is the link still working? & RQ$_4$\\
    Q10 & Is the implementation of the proposed solution publicly available? & RQ$_4$\\
    Q11 & Are the datasets used for training and/or evaluating the technique publicly available? & RQ$_4$\\\midrule
    Q12 & Do the researchers raise specific concern or discuss limitations about the experimented solutions? &  RQ$_5$\\
    \bottomrule
    \end{tabular}
    }
\end{table}

%%%%%%%%%%%%%%%%%%%%%%%%%%%%%%%%%%%%%%%%%
\subsection{Data Extraction and Analysis}
\label{sub:data_analysis}
%%%%%%%%%%%%%%%%%%%%%%%%%%%%%%%%%%%%%%%%%
The \selected primary studies have been inspected one last time with the goal of extracting the information needed to answer our RQs. The articles have been again equally split among the two authors with each of them in charge of extracting the needed data guided by the questionnaire in \tabref{tab:data}. The questions are clustered based on the RQ they serve.

Q1 collects the data needed to answer RQ$_1$ (\ie code review tasks automated in the literature). Q2-Q5 aim at categorizing the under-the-hood functioning of these techniques, thus answering RQ$_2$. Q6-Q8 shed light on the empirical evaluation performed to assess the proposed techniques (RQ$_3$) while Q9-Q11 look at the replicability of the primary studies and lists publicly available techniques and tools (RQ$_4$). Finally, Q12 informs our discussion of current limitations of automated techniques (RQ$_5$).

It is worth noting that some of the considered articles did not explicitly report some of the information we aim at collecting. Those cases are all documented in the \emph{master table} reported in our replication package \cite{replication}. 

Once collected the needed data, we answer our RQs as follows. For RQ$_1$ we report the list of code review tasks automated in the literature. Given this list, all other RQs are discussed by task. In all RQs in which ``categories'' must be defined (\eg the list of automated tasks in RQ$_1$), this has been obtained via an open-coding inspired procedure performed together by the two authors on the notes each of them took during the data extraction procedure, going back to the original paper if needed (\ie if the notes were not clear/comprehensive enough).

For RQ$_2$ we classify the automated approaches based on the technical solution they are built upon (\eg DL-based). Then, we distill findings about the training procedures followed for data-driven techniques and the targeted programming languages. For RQ$_3$ we focus instead on the evaluation, reporting the metrics usually adopted in the assessment of the techniques, whether qualitative analysis was present, and if the approach has been deployed in industry.

RQ$_4$ lists in a tabular fashion the available replication packages reporting for each of them whether they provide an implementation of the proposed technique and/or the datasets used in the study.

Finally, for RQ$_5$ we read the selected papers with a particular focus on the sections describing the approach, those discussing the results, and the conclusions to identify concerns/limitations about the proposed technique and its experimentation. We ignored classic limitations which can be found in any paper and which are usually discussed in the ``threats to validity'' section (\eg lack of generalizability beyond the scope of the experiment, limited hyperparameters tuning), but focused on concerns/limitations which are peculiar of the experimented technique (\eg the lack of an appropriate metric to assess its effectiveness). Once identified the relevant parts of the papers, a tag summarizing the discussed issue was defined. Then, similar tags were merged and the final list of tags was organized in a taxonomy presented in \tabref{tab:concerns}. The identified issues and their mapping with the corresponding papers were double-checked by a second author.
%!TEX root = main.tex
\newcommand{\rulec}{\arrayrulecolor{black}\specialrule{0.1em}{\abovetopsep}{0pt}}%
            
%%%%%%%%%%%%%%%%%
%%%%%%%%%%%%%%%%%
\section{Results} \label{sec:results}
%%%%%%%%%%%%%%%%%
%%%%%%%%%%%%%%%%%

\begin{figure*}[]
	\centering
	\includegraphics[width=0.7\linewidth]{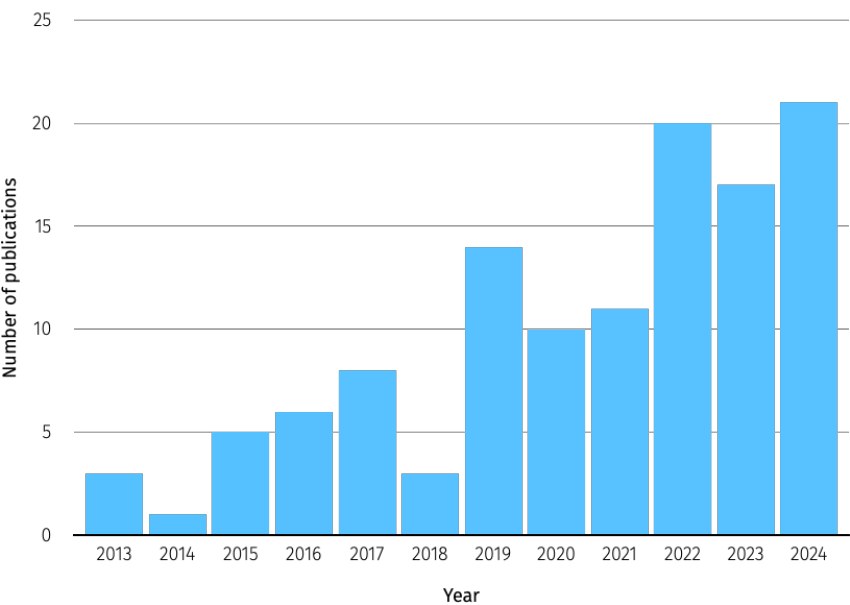}
	\caption{Publication years}
	\label{fig:years}
\end{figure*}

\begin{figure*}[]
	\centering
	\includegraphics[width=0.7\linewidth]{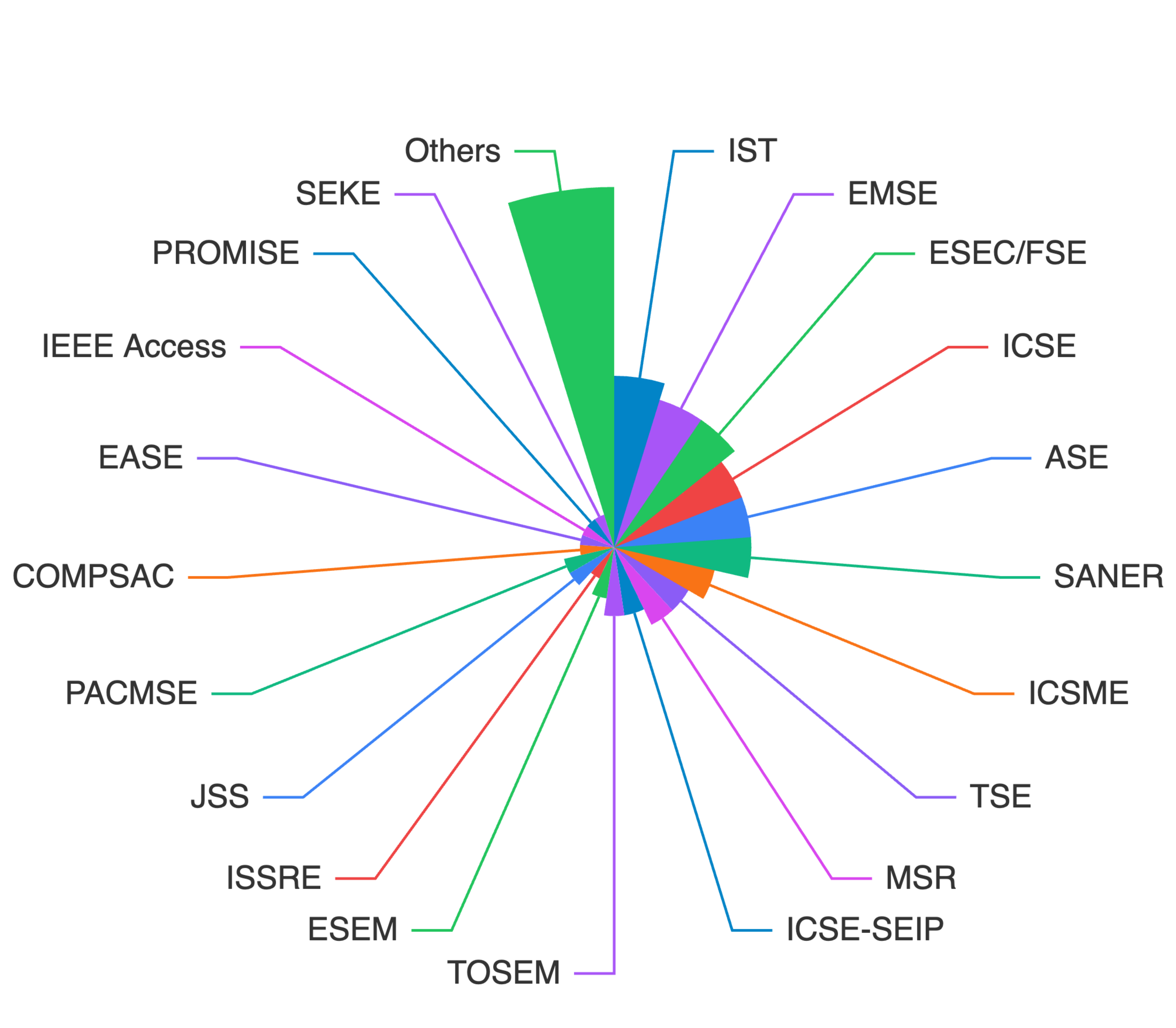}
	\caption{Publication venues}
	\label{fig:venues}
\end{figure*}

Before answering our RQs, we provide an overview about the identified primary studies. \figref{fig:years} plots the publication year of the \selected articles showing, with few exceptions, an overall increasing trend over the years with 21 papers published in 2024. \figref{fig:venues} shows instead the publication venues for these techniques, with venues such as IST, EMSE, ESEC/FSE, and ICSE being the most popular ones. \tabref{tab:venue_names} in the paper appendix indicates what each acronym used for publication venues stands for. 

\subsection{RQ$_1$: What are the code review tasks for which researchers proposed automated solutions?}

\tabref{tab:tasks} presents the \numberOfTasks types of code review tasks which have been automated in the literature. \tabref{tab:tasks} groups the tasks into macro categories (\eg ``Code Change Analysis'') and provides a short description of each task with related references (\ie the works addressing its automation). We discuss in the following each macro category.

\subsubsection{Assessing Review Quality} Works in this area aim at automatically assessing the quality of the review. Such information is  meant to be fed to the reviewer who can take proper actions to improve the review quality, if needed. 

Works in this area aimed at classifying review comments as useful or not-useful for the contributor \cite{pangsakulyanont:iwesep2014, rahman:msr2017, hasan:emse2021, yang:fse2023}. Rahman \etal \cite{rahman:esem2022} address a similar problem but by focusing specifically on comments requiring additional explanations to be properly understood by the contributor (thus being a subcategory of not-useful comments). Widyasari \etal \cite{widyasari:tosem2024} also investigate comments requiring additional explanations, also proposing the usage of Large Language Models (LLMs) to generate the additional explanations, when needed.

Finally, Hijazi \etal \cite{hijazi:tse2022, hijazi:issre2021} looked at the code review quality measurement from an orthogonal perspective using biometrics data. By monitoring the reviewer's activities (using \eg an eye-tracking device) they can provide feedback to the reviewer about areas of the reviewed code they did not pay enough attention to, thus suggesting a further check.

\subsubsection{Code Change Analysis} This category groups techniques aimed at analyzing the code change submitted for review in order to extract information useful to support the reviewer in its inspection. Several authors \cite{tao:msr2015, barnett:icse2015, wang:ase2019} targeted the splitting of tangled commits \cite{Herzig:msr2013} into smaller and cohesive changes which are supposed to be easier to review. Indeed, having smaller changes can help in achieving quick review turnarounds \cite{bacchelli:icse2013,sadowski:icse2018} while cohesive changes simplify the identification of proper reviewers, which are more likely to have a comprehensive expertise to review the change (given its cohesiveness and focus).

Huang \etal \cite{huang:tse2020} propose the automated identification of the ``salient-class'' in a commit to review. The salient-class is the one supposed to be the main focus of the changes and which likely triggered changes to other code locations. Such a class can be used as entry point for the review process, assuming that this will simplify the code change understanding. 

Wang \etal \cite{wang:ist2021a} suggest the automated linking of similar contributions which may help in identifying duplicated patches and, more in general, in increasing the reviewers' awareness about changes impacting similar locations, thus promoting a better code review. 

Finally, with the goal of minimizing the number of code review iterations needed to accept a proposed change, Hong \etal \cite{hong:ist2024} propose a change impact analysis methodology specifically tailored for the code review process and aimed at identifying functions that must co-change given the proposed contribution, but are not changed.

\subsubsection{Code Change Classification} Works in this area classify the whole code change to review again with the goal of augmenting the information available to reviewers before starting the code inspection. Predicting whether the code change will be approved (merged) or needs additional review rounds is the most popular code change classification task tackled in the literature \cite{lu:issre2023, nashaat:tse2024, wu:kbs2022, wu2022contrastive, li:fse2022, shi:2019, fan:emse2018, islam:ist2022, chouchen:tosem2024, yang:emse2024, chouchen:emse2024}. Works on this topic provide a representation of the code change as input to the approach (\eg to a DL model) expecting it to suggest whether the implemented change is acceptable. A variation is to also provide the technique with information about the specific change the developer was asked to implement (\eg a reviewer comment that the contributor had to address). The outputted boolean prediction can help, for example, to prioritize the diff hunks part of a pull request, focusing on those likely to require a reviewer's comment \cite{li:fse2022}.

%!TEX root = ../main.tex

%%%%%%%%%%%%%%%%%
%%%%%%%%%%%%%%%%%
%%%%%  Table Tasks  %%%%%
%%%%%%%%%%%%%%%%%
%%%%%%%%%%%%%%%%%

\begin{table}
	\centering
    \caption{Code review tasks for which automated solutions have been proposed\vspace{-0.3cm}}
    \label{tab:tasks}
    {\footnotesize
   \begin{tabular}{ p{0.00001cm} p{0.00001cm} | p{4.7cm} | p{6.9cm} | p{1.5cm}} 
    \rulec
    \multicolumn{2}{c|}{\bf Type} & {\bf Task} & {\bf Description} & {\bf Reference}\\
    \rulec
    
  % ASSESSING REVIEW QUALITY
  
  \multirow{6}{*}{\rotatebox[origin=c]{90}{Assessing}} & \multirow{6}{*}{\rotatebox[origin=c]{90}{Review Quality}}  
&\multirow{2}{*}{Assessing Review Quality through Biometrics} 
&Evaluate the quality of code review using biometrics data, warning the reviewer if specific areas of code deserve a further check 
&\multirow{2}{*}{\centering  \cite{hijazi:tse2022, hijazi:issre2021}} \\ %[0.05cm]

& &\cellcolor{lightlightgrey}\multirow{2}{*}{Classifying the Usefulness of Review Comments} 
&\cellcolor{lightlightgrey}Classify a given code review comment as useful or not-useful for the contributor 
&\multirow{2}{*}{\centering \cellcolor{lightlightgrey}\cite{pangsakulyanont:iwesep2014, rahman:msr2017, hasan:emse2021, yang:fse2023}} \\ %[0.05cm]

& &Identifying/Improving Review Comments Needing Further Explanations 
&Identifies review comments which need further explanations to be properly understood by the contributor 
&\multirow{2}{*}{\centering  \cite{rahman:esem2022, widyasari:tosem2024}}\\ %[0.05cm]
% & &\multirow{2}{*}{Estimate Defect Content After a Review} & Assess the likelihood of a change to featured defects after the code review &\cite{wohlin:icse1998, petersson:issre2001} \\[0.05cm]
\hline

% CODE CHANGE ANALYSIS
\multirow{7}{*}{\rotatebox[origin=c]{90}{Code Change}} & \multirow{7}{*}{\rotatebox[origin=c]{90}{Analysis}}  
&\cellcolor{lightlightgrey}Decomposing Tangled Commit 
&\cellcolor{lightlightgrey}Split a composite code change into smaller and cohesive changes 
&\cellcolor{lightlightgrey}\cite{tao:msr2015, barnett:icse2015, wang:ase2019} \\ %[0.05cm] 

&  &\multirow{2}{*}{Impact Analysis for Code Review} 
&Recommend functions that must be changed given the submitted contribution 
&\multirow{2}{*}{\cite{hong:ist2024}}\\ %[0.05cm] 

&  &\multirow{2}{*}{\cellcolor{lightlightgrey}Linking Similar Contributions} 
&\cellcolor{lightlightgrey}Link similar changes to review that share textual content and modify similar code locations 
&\multirow{2}{*}{\cellcolor{lightlightgrey}\cite{wang:ist2021a, ayinala:compsac2020}}\\ 

&  &\multirow{2}{*}{Predicting Salient-Class} 
&Identification of the “salient-class” in a commit to review, namely the class causing the other changes in the commit 
&\multirow{2}{*}{\cite{huang:tse2020}}\\ 
\hline

 % CODE CHANGE CLASSIFICATION
 \multirow{6}{*}{\rotatebox[origin=c]{90}{Code Change}} & \multirow{6}{*}{\rotatebox[origin=c]{90}{Classification}}  
&\cellcolor{lightlightgrey}Identifying Impactful Code Changes 
&\cellcolor{lightlightgrey}Identify impactful code changes (\eg impacting the system design) 
&\cellcolor{lightlightgrey}\cite{wen:icsme2018, uchoa:msr2021} \\ 

&  &Identifying Large-review-effort Code Changes 
&Identify code changes that will require a large reviewing effort 
&\cite{wang:ist2021b}\\ 
  %&  &\cellcolor{lightlightgrey}Classifying the Type of Code Change &\cellcolor{lightlightgrey} Classify the scope of a code change (\eg functional, evolvability) &\cellcolor{lightlightgrey}\cite{fregnan:emse2022} \\[0.05cm]

&  &\cellcolor{lightlightgrey}\multirow{2}{*}{Identifying Quickly Reviewable Changes} 
&\cellcolor{lightlightgrey}Rank changes to be reviewed based on their likelihood of being quickly merged or rejected 
&\multirow{2}{*}{\cellcolor{lightlightgrey}\cite{zhao:emse2019}}\\ 

& &Predicting Code Changes Approval, Merge, or Need for review 
&Predict the likelihood of a change of being accepted, merged, or needing review 
&\cite{lu:issre2023, nashaat:tse2024, wu:kbs2022, wu2022contrastive, li:fse2022, shi:2019, fan:emse2018, islam:ist2022, chouchen:tosem2024, yang:emse2024, chouchen:emse2024} \\ 
\hline

% CODE CHANGE QUALITY CHECK
\multirow{10}{*}{\rotatebox[origin=c]{90}{Code Change}} & \multirow{10}{*}{\rotatebox[origin=c]{90}{Quality Check}}  
&\cellcolor{lightlightgrey}\multirow{2}{*}{Checking Design Patterns Consistency} 
&\cellcolor{lightlightgrey} Check whether the implemented change violates existing design patterns 
&\multirow{2}{*}{\cellcolor{lightlightgrey}\cite{he:icssrc2013}}\\ 

& &Generating Review Comments
&Generate review comments for a given piece of code
&\cite{li.l:esecfse2022, li:fse2022, hong:esecfse2022, tufano:icse2022, vijayvergiya:aiware2024, lu:ase2024, yu:tosem2024, lin:msr2024, sghaier:fse2024, lu:issre2023, nashaat:tse2024}\\ 

& &\cellcolor{lightlightgrey}Identifying Clone Refactoring Opportunities 
&\cellcolor{lightlightgrey}Detect unrefactored or partially refactored code clones 
&\cellcolor{lightlightgrey}\cite{chen:compsac2017}\\ 

& &Predicting Code Defectiveness 
&Predict the defectiveness of a patch before or after being reviewed 
&\cite{soltanifar:esem2016, sharma:spe2019} \\ 

& &\multirow{2}{*}{\cellcolor{lightlightgrey}Predicting Problematic Code Elements} 
&\cellcolor{lightlightgrey}Predict code elements in a given contribution reviewers should pay particular attention to (\eg lines likely needing changes) 
&\multirow{2}{*}{\cellcolor{lightlightgrey}\cite{hong:saner2022, sghaier:saner2023, olewicki:fse2024}}\\ 

& &Reviewing Code Formatting Violations 
&Suggest how to fix code formatting violations in a given piece of code 
&\cite{markovtsev:msr2019}\\

& &\cellcolor{lightlightgrey}Reviewing via Static Analysis 
&\cellcolor{lightlightgrey}Use multiple static analysis tools to generate a code review 
&\cellcolor{lightlightgrey}\cite{balachandran:icse2013}\\ 

\hline
 
% CODE REVIEW SENTIMENT ANALYSIS
\multirow{7}{*}{\rotatebox[origin=c]{90}{Code Review}} & \multirow{7}{*}{\rotatebox[origin=c]{90}{Sent. Analysis}} 
&\multirow{2}{*}{Classifying the Sentiment of Review Comments}
&Classify the sentiment of  review comments as neutral, negative, or positive
&\multirow{2}{*}{\cite{ahmed:ase2017}}\\

& &\multirow{2}{*}{\cellcolor{lightlightgrey}Identifying “Pushback” Feelings in Reviews} 
&\cellcolor{lightlightgrey}Identify feelings of “pushback", with the reviewer blocking a change request for interpersonal conflicts 
&\multirow{2}{*}{\cellcolor{lightlightgrey}\cite{egelman:icse2020}}\\ 

& &Identifying Toxic/Uncivil Review Comments 
&Identify toxic or uncivil comments in code reviews 
&\cite{sarker:tosem2023, ferreira:jss2024, sarker:esem2023, rahman:fse2024} \\

& &\multirow{2}{*}{\cellcolor{lightlightgrey}Rephrasing Toxic/Uncivil Comments} 
&\cellcolor{lightlightgrey}Rephrase review comments to improve its politeness without changing its semantic 
&\multirow{2}{*}{\cellcolor{lightlightgrey}\cite{rahman:fse2024}} \\ 
\hline
  
% RETRIEVAL OF SIMILAR CR/CC
\multirow{6}{*}{\rotatebox[origin=c]{90}{Retrieval of}} & \multirow{6}{*}{\rotatebox[origin=c]{90}{Similar CR/CC}} 
&\multirow{4}{*}{Augmenting Reviews} 
& Can be used to provide either (i) the contributor with examples of reviews similar to those they are receiving (for better understanding); or (ii) the reviewer with examples of reviews which have been written for code similar to the one they are inspecting 
&\multirow{4}{*}{\cite{gupta:sigkdd2018, guo2020review, siow:saner2020, rahman:esem2022, shuvo:icsme2023, kartal:infsof2024, guo:saner2019, hirao:esecfse2019}}\\

& &\multirow{2}{*}{\cellcolor{lightlightgrey}Mining Code Improvement Patterns} 
&\cellcolor{lightlightgrey}Extract source code improvement patterns from existing code review history to recommend how to improve the submitted code 
&\multirow{2}{*}{\cellcolor{lightlightgrey}\cite{ueda:iwsc2019}} \\
\hline
 
% REVISED CODE GENERATION
\multirow{5}{*}{\rotatebox[origin=c]{90}{Revised Code}} & \multirow{5}{*}{\rotatebox[origin=c]{90}{Generation}} 
&\multirow{3}{*}{\parbox{4.7cm}{Implementing the Code Change Requested by a Reviewer}}
&Generate a revised version of a given piece of code by implementing a specific change requested by the reviewer in a natural language comment 
&\multirow{3}{*}{\parbox{1.5cm}{\cite{tufano:icse2021, tufano:icse2022, huq:ist2022, li:fse2022, sghaier:fse2024, lu:issre2023, nashaat:tse2024, zhang:ase2022, pornprasit:ist2024, lu:apsec2023, froemmgen:icse2024}}}\\

& &\cellcolor{lightlightgrey}Predicting the Code Output of the Review Process 
&\cellcolor{lightlightgrey}Given a code snippet submitted for review, revise it to implement changes which are likely to be required by reviewers 
&\multirow{2}{*}{\cellcolor{lightlightgrey}\cite{tufano:icse2021, tufano:icse2022, patanamon:icse2022, pornprasit:saner2023}}\\
\hline
  
 % TIME MANAGEMENT
\multirow{6}{*}{\rotatebox[origin=c]{90}{Time}} & \multirow{6}{*}{\rotatebox[origin=c]{90}{Management}} 
&\multirow{2}{*}{Identifying Blocking Actors in Pull Requests} 
&Identify who among contributor(s) and reviewer(s) is to blame for overdue pull requests
&\multirow{2}{*}{\cite{shan:esecfse2022}}\\

& &\cellcolor{lightlightgrey}Predicting Pull Request/Code Review Completion Time 
&\multirow{2}{*}{\cellcolor{lightlightgrey}Predict the time needed to complete a pull requests/code review} 
&\multirow{2}{*}{\cellcolor{lightlightgrey}\cite{maddila:esefse2019, shan:esecfse2022, chouchen:emse2023, chen:esecfse2022,yang:emse2024}}\\

% & &\cellcolor{lightlightgrey}\multirow{2}{*}{Identifying Overdue Pull Requests} &\cellcolor{lightlightgrey}Identify overdue pull requests (i.e., pull requests taking longer than the expected resolution time) &\cellcolor{lightlightgrey}\cite{shan:esecfse2022}\\[0.05cm]

& &\multirow{2}{*}{Prioritizing Review Requests} 
&Prioritize code review requests based on factors such as age of the change, test verdicts, etc. 
&\multirow{2}{*}{\cite{saini:icse-seip2021, chouchen:emse2024, yang:ase2024}}\\
\hline
  
 % OTHER
\multicolumn{2}{c|}{\multirow{10}{*}{\rotatebox[origin=c]{90}{Other}}}
&\cellcolor{lightlightgrey}\multirow{3}{*}{\parbox{4.7cm}{\centering Classifying the Goal of a Review Comment or the Type of Change Triggered by a Comment}}
&\cellcolor{lightlightgrey}Classify a review comment as Style, Functionality, Test, Approval, Disagreeing, Questioning, Roadmap, Diversion, Convention, Response or Encouragement 
&\multirow{3}{*}{\cellcolor{lightlightgrey}\cite{li:seke2017, fregnan:emse2022, Turzo:esem2023}}\\

& &\multirow{2}{*}{Configuring Static Code Analysis Tools} 
&Leverage code review comments for recommending static code analysis tools and warning categories to be used in future 
&\multirow{2}{*}{\cite{zampetti:emse2022}} \\

%& &Generating Review Checklist &Generate a checklist to guide the reviewer’s inspection &\cite{belli:issre1996}\\[0.05cm]

& &\multirow{2}{*}{\cellcolor{lightlightgrey}Partitioning Static Analysis Warnings}
&\cellcolor{lightlightgrey}Cluster the warnings of static analysis tools into categories to simplify their inspection 
&\multirow{2}{*}{\cellcolor{lightlightgrey}\cite{tukaram:scam2013}}\\
 
& &Recommending Reviewers 
&Recommend reviewers that are best suited for the given piece of code 
&\cite{balachandran:icse2013, jiang:jcst2015, thongtanunam:saner2015, xia:icsme2015, ouni:icsme2016, ying:csi2016, yu:ist2016, zanjan:tse2016, xia:sm2017,  jiang:ist2017, fejzer:jiis2018, asthana:esecfse2019, liao:globecom2019, sulun:icpmdase2019, jiang:jss2019, mirsaeedi:icse2020, al:icpmda2020, strand:icse-sep2020, chouchen:asc2021, tecimer:ease2021, pandya:esecfse2022, li:ease2023, aryendu:ase2023, zhang:icse2023, rahman:icse2016, chueshev:icsme2020, rebai:ase2020, ye:ieee2019, zhao:cascon:2022, qiao:saner2024, hajari:tse2024, rahman:icsme2023, rong:icse2022, sulun:ist2021, kong:saner2022, ahasanuzzaman:emse2024}\\

& &\multirow{2}{*}{\cellcolor{lightlightgrey}Visualizing Code Changes} 
&\cellcolor{lightlightgrey}Provide visualizations of the change to review to ease code comprehension 
&\multirow{2}{*}{\cellcolor{lightlightgrey}\cite{menarini:ase2017, fadhel:iccq2021, brito:icpc2021, fregnan:jss2023}}\\

    \rulec
    \end{tabular}
    }
\end{table}

Another line of research aims at identifying code contributions which, due to their nature, will require a large review effort. Uch\^oa \etal \cite{uchoa:msr2021} automatically flag code changes which are likely to impact the software design, thus requiring extra care in their assessment.  Wen \etal \cite{wen:icsme2018} propose BLIMP Tracer, a tool to support code review through impact analysis information, thus helping in identifying changes impacting mission-critical deliverables. Wang \etal \cite{wang:ist2021b} generalize the problem to the automated identification of large-review-effort changes while, at the other side of the spectrum, Zhao \etal \cite{zhao:emse2019} target the identification of quickly reviewable changes, namely contributions that are easy to merge or reject. Similarly to the work classifying the contributions as likely to be accepted/rejected, all these works provide code reviewers with information useful for prioritizing the changes to inspect. 

\subsubsection{Code Change Quality Check} Researchers proposed solutions to (partially) automate the quality check usually in place when reviewing a code change. Approaches addressing this task substantially vary in their goal and complexity. Some of them focus on specific code quality aspects, such as predicting whether a submitted patch is likely to introduce a bug \cite{soltanifar:esem2016, sharma:spe2019}, identifying the presence of missed clone refactoring opportunities \cite{chen:compsac2017}, or checking whether the implemented change violates existing design patterns \cite{he:icssrc2013}. Other techniques address the same problem with, however, a more general view on code quality. 

Some authors \cite{hong:saner2022, sghaier:saner2023, olewicki:fse2024} aim at predicting code elements in a patch which require the reviewer's attention, since likely in need for changes. These approaches are useful in the context of within-patch review prioritization (\ie deciding where to allocate more review effort within a patch). Other works push the boundaries further targeting the automated generation of concrete feedback for the contributor, as a human reviewer would do. A first strategy to achieve this goal consists in merging the output of several static analysis tools \cite{balachandran:icse2013}, providing the contributor with a list of potential flaws identified in the submitted patch. The most recent trend consists, however, in exploiting DL models to generate natural language comments for a given patch, with the model imitating a human reviewer \cite{li.l:esecfse2022, li:fse2022, hong:esecfse2022, tufano:icse2022, vijayvergiya:aiware2024, lu:ase2024, yu:tosem2024, lin:msr2024, sghaier:fse2024, lu:issre2023, nashaat:tse2024}. These techniques are trained on thousands of examples of real code reviews (\ie review comments liked to specific code changes) and can then be applied to previously unseen changes to generate review comments. Markovtsev \etal \cite{markovtsev:msr2019} focused on a simplified version of this problem: Their approach ``learns'' the code formatting style of a given software project, identifies violations to such a style, and suggests possibly fixes as automatically generated reviewer's comments.

\subsubsection{Code Review Sentiment Analysis} The code review process may result in critiques moved by a developer (reviewer) to one of their peers (contributor). The way in which these critiques are formalized in the reviewer's comment can play an important role in the successful outcome of the whole process. For this reason, researchers applied sentiment analysis techniques to automatically classify the sentiment of reviewers' comments \cite{ahmed:ase2017}: Flagging comments expressing a negative sentiment can provide useful information to the reviewer, who can revise those potentially problematic comments. 

Other authors tackled a more specific version of this problem, focusing on the identification of a specific type of reviewers' comments expressing negative feelings. In particular, Egelman \etal \cite{egelman:icse2020} aim at identify review comments suggesting the will of the reviewer to block a change request for interpersonal conflicts rather than for the quality of the submitted contribution. Sarker \etal \cite{sarker:esem2023,sarker:tosem2023}, instead, focus on the identification of ``toxic code reviews'', while Ferreira \etal \cite{ferreira:jss2024} and Rahman \etal \cite{rahman:fse2024} target ``uncivil review comments''. Incivility represents a broader set of negative comments as compared to \emph{toxicity}, since the latter entails hate speech and offensive language, while incivility does not \cite{ferreira:jss2024}. Note that Rahman \etal \cite{rahman:fse2024}, besides identifying uncivil comments, also present a model able to propose alternative civil rephrasing preserving the original comments' semantic.

\subsubsection{Retrieval of Similar Code Reviews/Code Changes} Retrieval techniques have been used to create recommender systems supporting code review from different perspectives. Given a code fragment to review, some techniques \cite{gupta:sigkdd2018, guo2020review, siow:saner2020, rahman:esem2022, shuvo:icsme2023, kartal:infsof2024, guo:saner2019, hirao:esecfse2019} retrieve from a dataset of past reviews those involving similar code fragments and recommend to the reviewer comments they can reuse (since used in the past to suggest improvements to similar code). Rahman \etal \cite{rahman:esem2022} also proposed a similar approach, but motivated it as a mechanism to provide the contributor with additional examples of reviews similar to those they are receiving. This could help in better understanding what the reviewer meant.

Ueda \etal \cite{ueda:iwsc2019} focused instead on mining recurring improvement patterns from code review (\ie changes frequently suggested by reviewers). Those patterns can then be potentially applied to improve the quality of the code to review (even before the review process starts).

\subsubsection{Revised Code Generation} This line of research aims at supporting the code review process by automatically generating the code output of the review process. Two variations of this task have been proposed. The fist \cite{tufano:icse2021, tufano:icse2022, patanamon:icse2022, pornprasit:saner2023} provides as input to the automated technique a code snippet submitted for review and expects the technique to revise such a code to implement changes which will likely be requested during the code review process. These techniques are meant to be used by the contributor before even starting the code review process to quickly verify whether improvements can be made to the code they write. 

The second \cite{tufano:icse2021, tufano:icse2022, huq:ist2022, li:fse2022, sghaier:fse2024, lu:issre2023, nashaat:tse2024, zhang:ase2022, pornprasit:ist2024, lu:apsec2023, froemmgen:icse2024} is instead a \emph{code refinement} task in which the approaches are provided as input not only a code snippet submitted for review but also a specific reviewer's comment to address. In this case the goal of the approach is to automatically revise the submitted code generating a version of it addressing the comment provided as input. These approaches are meant to be used during the code review process either (i) by the reviewer, to attach to their comments an example of how they envision the revised code, or (ii) by the contributor, to automatically address some of the reviewer's requests.

\subsubsection{Time Management} Evidence from the literature suggests that both open source and industrial projects can undergo hundreds of reviews per month (\eg $\sim$500 reviews per month in Linux \cite{Rigby:tosem2014}, $\sim$3k in Microsoft Bing \cite{Rigby:fse2013}). In such a context time management becomes essential and researchers proposed solutions to help the proper allocation of reviewers' time. Differently from previously discussed techniques which automated specific code review tasks, these approaches aim at augmenting the information available to reviewers and/or managers, thus possibly improving decisions taken during code review.

Some of the proposed solutions can be combined in a sort of pipeline to support the code review: Approaches to predict the time needed to complete a pull request \cite{maddila:esefse2019, shan:esecfse2022, chouchen:emse2023, chen:esecfse2022,yang:emse2024} can be used to inform techniques aimed at prioritizing review requests \cite{saini:icse-seip2021, chouchen:emse2024, yang:ase2024}. Also, pull requests taking longer than expected can be provided as input to techniques identifying blocking actor(s) \cite{shan:esecfse2022}, namely the person(s) responsible for the delay. This could help in triggering the blocking actor or, if possible, replace them.

%Still with the goal of optimizing the review time, Saini \etal \cite{saini:icse-seip2021} presented a technique to prioritize code review requests considering factors such as the age of the change, whether the change passed or not the tests, the number of revisions that have been already done for that change, \etc

\subsubsection{Other} The last category groups together tasks which did not fit in the previously presented categories and features heterogeneous tasks. These include the code review task which has been mostly subject to automation attempts in the literature: the recommendation of reviewers that are best suited for a given change \cite{balachandran:icse2013, jiang:jcst2015, thongtanunam:saner2015, xia:icsme2015, ouni:icsme2016, ying:csi2016, yu:ist2016, zanjan:tse2016, xia:sm2017,  jiang:ist2017, fejzer:jiis2018, asthana:esecfse2019, liao:globecom2019, sulun:icpmdase2019, jiang:jss2019, mirsaeedi:icse2020, al:icpmda2020, strand:icse-sep2020, chouchen:asc2021, tecimer:ease2021, pandya:esecfse2022, li:ease2023, aryendu:ase2023, zhang:icse2023, rahman:icse2016, chueshev:icsme2020, rebai:ase2020, ye:ieee2019, zhao:cascon:2022, qiao:saner2024, hajari:tse2024, rahman:icsme2023, rong:icse2022, sulun:ist2021, kong:saner2022, ahasanuzzaman:emse2024}. These techniques, while sharing the same goal, differ for the underlying technical solution adopted (RQ$_2$ focuses on this aspect) and for the features used to rank the reviewers given the change. In most of cases the features include information extracted from the history of code changes to favor the recommendation of reviewers who \eg already worked in the past on the code files subject of the change or already reviewed similar patches. The recency of these activities is usually considered as well.

Another popular task in the ``Other'' category features approaches providing visualizations for the code changes to review in order to simplify the reviewer's inspection \cite{menarini:ase2017, fadhel:iccq2021, brito:icpc2021, fregnan:jss2023}. Note that we only included in our SLR visualization techniques specifically aimed at supporting code review. Different works focus the visualization on different types of information. Brito and Valente \cite{brito:icpc2021} propose RAID, a tool for refactoring-aware code review which visualizes the refactoring operations implemented in the change to review. Fadhel and Sekerinski \cite{fadhel:iccq2021} target instead visualizations aimed at improving the reviewer's awareness of the possible impact that the implemented changes can have on the system's architecture. Fregnan \etal \cite{fregnan:jss2023} provide a more general-purpose graph-based visualization to support code review: Each node represents a class or a method and the links between them represents dependencies such as method calls. The goal here is to improve the navigation of the change and its comprehension. Finally, still related to visualization is the behavioral diff generated by the approach proposed in \cite{menarini:ase2017}. The idea is to show the behavioral differences (in terms of test case execution) which can be observed in the system before and after the implementation of the code change to review. This can support the assessment of code change correctness made by the reviewer.

Moving to the next task, Li \etal \cite{li:seke2017} present an approach to automatically classify reviewers' comments into the categories reported in \tabref{tab:tasks} (\eg style, functionality, \etc). Their approach is meant to provide a better understanding and monitoring of the ongoing review process. On top of that, with the proposal of data-driven techniques to automate tasks such as \emph{generating review comments} this approach can be used to cleanup the training set of these techniques, removing for example the comments classified as ``Encouragement'', since irrelevant for training techniques suggesting how to improve code snippets. A similar approach has also been presented by Turzo \etal \cite{Turzo:esem2023}, while Fregnan \etal \cite{fregnan:emse2022} focus on classifying the code changes implemented as result of the code review process.

Tukaram \etal \cite{tukaram:scam2013} propose the idea of partitioning static analysis warnings, with the goal of clustering the similar ones thus simplifying their interpretation. On a related research thread, Zampetti \etal \cite{zampetti:emse2022} suggest the automated analysis of review comments posted in the past to understand which static analysis tools should be used in the continuous integration pipeline of a given project and how they should be configured. In other words, they aim at understanding what the relevant ``issues'' reviewers look for when inspecting a patch and which of those issues can be automatically identified by static analysis tools.

\pgfplotsset{testbar/.style={
width=.2\textwidth,
xbar stacked,
axis y line= none, axis x line= none,
xmajorgrids = false,
xmin=0, xmax=100,
bar width=1.5mm, y=1mm
}}

%!TEX root = ../main.tex

% LEGEND APPROACH
% \addplot[fill=cyan!40] coordinates{(0,0) };       % DL
% \addplot[fill=red!40] coordinates{(0,0)};          % ML
% \addplot[fill=violet!40] coordinates{(0,0) };     % IR
% \addplot[fill=orange!40] coordinates{(0,0) };   % HB
% \addplot[fill=gray!40] coordinates{(0,0) };       % Other

% LEGEND LANGUAGE
% \addplot[fill=pink!40] coordinates{(0,0) };      % Java
% \addplot[fill=yellow!40] coordinates{(0,0) };   % Multiple Languages
% \addplot[fill=purple!40] coordinates{(0,0) };   % Language Independent
% \addplot[fill=gray!40] coordinates{(0,0) };     % Other

%% `>{\arraybackslash\centering}` <--- used to center horizzontally in the cell
%% `\raisebox{-0.5\height}{\strut%` <--- used to center vertically in the cell
%% '\makebox[0pt][l]{\cellcolor{lightlightgrey}}%' <--- used to center horizontally the Tikz rectangle when coloring the cell

\newcommand{\cbox}[1]{\raisebox{0.1cm}{\fcolorbox{black}{#1}{\null}}}

\begin{table}
\centering
\captionsetup{justification=centering}
\caption{
Under-the-hood solutions behind the techniques and tools proposed for code review automation.\\
Approaches: \cbox{cyan!40} Deep Learning; \cbox{red!40} Machine Learning; \cbox{violet!40} Information Retrivial; \cbox{orange!40} Heuristic-Based; \cbox{gray!40} Other.\\
Programming Languages: \cbox{pink!40} Java; \cbox{yellow!40} Multiple Languages; \cbox{purple!40} Language Independent; \cbox{gray!40} Other. \vspace{-0.2cm}}

\label{tab:rq2}
    {\scriptsize
   \begin{tabular}{ m{4.5cm} | c | >{\arraybackslash\centering}p{0.68cm}| >{\arraybackslash\centering}p{0.68cm}| >{\arraybackslash\centering}p{1cm}| p{2.6cm} | c}
    \rulec
  \multirow{2}{*}{\bf Task} & \multirow{2}{*}{\bf Approach} &\multicolumn{3}{c|}{\bf Training} & \multirow{2}{*}{\bf Granularity}&\multirow{2}{*}{\bf Language}\\\cline{3-5}
                                                               &                    &PT NL&PT code&FT&               &\\
    \rulec
   %  & \\[-0.2cm]
   
    %ASSESSING REVIEW QUALITY
      \multicolumn{7}{c}{\bf  Assessing Review Quality} \\ \rulec
    % ------------------------------------------------------------------------------------------------------------ end row
     
    \raisebox{-0.3\height}{\strut Assessing Review Quality through Biometrics (2)}&
      
    \raisebox{-0.5\height}{\strut% 
    \begin{tikzpicture}
    \begin{axis}[testbar] 
    \addplot[fill=red!40] coordinates{(100,0)}; %ML
    \end{axis}
    \end{tikzpicture}}
    
     & 
     \raisebox{-0.4\height}{\strut \xmark} & 
     \raisebox{-0.4\height}{\strut \xmark} &  
     \raisebox{-0.4\height}{\strut \xmark}  & 
     \raisebox{-0.4\height}{\strut code regions, code review} & 
     
    \raisebox{-0.5\height}{\strut%
    \begin{tikzpicture}
    \begin{axis}[testbar] 
    \addplot[fill=gray!40] coordinates{(100,0) }; %other C
    \end{axis}
    \end{tikzpicture}}
     \\  
     % ------------------------------------------------------------------------------------------------------------ end row
     
    \cellcolor{lightlightgrey} \raisebox{-0.3\height}{\strut Classifying the Usefulness of Review Comments (4)}& 
     
     \makebox[0pt][l]{\cellcolor{lightlightgrey}}%
    \raisebox{-0.5\height}{\strut%
    \begin{tikzpicture}
    \begin{axis}[testbar] 
    \addplot[fill=cyan!40] coordinates{(25,0) };      % DL 1/4
    \addplot[fill=red!40] coordinates{(50,0)};        %ML 2/4
    \addplot[fill=orange!40] coordinates{(25,0) }; %HB 1/4
    \end{axis}
    \end{tikzpicture}}
    
     & 
      \cellcolor{lightlightgrey}\raisebox{-0.4\height}{\strut 1/4} & 
      \cellcolor{lightlightgrey}\raisebox{-0.4\height}{\strut \xmark} & 
      \cellcolor{lightlightgrey}\raisebox{-0.4\height}{\strut 4/4} & 
      \cellcolor{lightlightgrey}\raisebox{-0.4\height}{\strut review comment} & 
     
    \makebox[0pt][l]{\cellcolor{lightlightgrey}}%
    \raisebox{-0.5\height}{\strut% 
    \begin{tikzpicture}
    \begin{axis}[testbar] 
    \addplot[fill=purple!40] coordinates{(75,0) }; % language independent
    \addplot[fill=gray!40] coordinates{(25,0) }; % Other (python)
    \end{axis}
    \end{tikzpicture}}
     \\ 
     % ------------------------------------------------------------------------------------------------------------ end row

    Identifying/Improving Review Comments Needing Further Explanations (2)& 
 
    \raisebox{-0.3\height}{\strut%
    \begin{tikzpicture}
    \begin{axis}[testbar] 
    \addplot[fill=cyan!40] coordinates{(50,0) }; % DL
    \addplot[fill=gray!40] coordinates{(50,0)};  %ML
    \end{axis}
    \end{tikzpicture}}
    
     & 
     \raisebox{-0.1\height}{\strut 1/2} & 
     \raisebox{-0.1\height}{\strut 1/2} &  
     \raisebox{-0.1\height}{\strut 1/2 } & 
     \raisebox{-0.1\height}{\strut review comment} & 
     
    \raisebox{-0.3\height}{\strut%
    \begin{tikzpicture}
    \begin{axis}[testbar] 
    \addplot[fill=purple!40] coordinates{(100,0) }; % language independent
    \end{axis}
    \end{tikzpicture}}\\ 
     \rulec
     % ------------------------------------------------------------------------------------------------------------ end row
    
    % CODE CHANGE ANALYSIS
  \multicolumn{7}{c}{\bf Code Change Analysis} \\ \rulec
    % ------------------------------------------------------------------------------------------------------------ end row
  
      \cellcolor{lightlightgrey}\raisebox{-0.3\height}{\strut Decomposing Tangled Commit (3)}&
     
    \makebox[0pt][l]{\cellcolor{lightlightgrey}}%
    \raisebox{-0.5\height}{\strut%
    \begin{tikzpicture}
    \begin{axis}[testbar] 
    \addplot[fill=orange!40] coordinates{(100,0) }; %HB
    \end{axis}
    \end{tikzpicture}}
    
     & 
     \cellcolor{lightlightgrey}\raisebox{-0.4\height}{\strut \xmark} & 
     \cellcolor{lightlightgrey}\raisebox{-0.4\height}{\strut \xmark} & 
     \cellcolor{lightlightgrey}\raisebox{-0.4\height}{\strut \xmark}  & 
     \cellcolor{lightlightgrey}\raisebox{-0.4\height}{\strut commit} & 
     
     \makebox[0pt][l]{\cellcolor{lightlightgrey}}% 
     \raisebox{-0.5\height}{\strut%
     \begin{tikzpicture}
    \begin{axis}[testbar] 
    \addplot[fill=pink!40] coordinates{(66.5,0) }; % Java
    \addplot[fill=gray!40] coordinates{(33.5,0) }; % Other
    \end{axis}
    \end{tikzpicture}} \\
     % ------------------------------------------------------------------------------------------------------------ end row
     
   \raisebox{-0.3\height}{\strut Impact Analysis for Code Review (1)}&
     
    \raisebox{-0.5\height}{\strut%
    \begin{tikzpicture}
    \begin{axis}[testbar] 
    \addplot[fill=gray!40] coordinates{(100,0) };       % Other
    \end{axis}
    \end{tikzpicture}}
    
    &
    \raisebox{-0.4\height}{\strut \xmark} &
    \raisebox{-0.4\height}{\strut \xmark} &
    \raisebox{-0.4\height}{\strut \xmark} &
    \raisebox{-0.4\height}{\strut PR} & 
    
    \raisebox{-0.5\height}{\strut%
    \begin{tikzpicture}
    \begin{axis}[testbar] 
    \addplot[fill=purple!40] coordinates{(100,0) };   % Language Independent
    \end{axis}
    \end{tikzpicture}}\\
    % ------------------------------------------------------------------------------------------------------------ end row
    
    \cellcolor{lightlightgrey}\raisebox{-0.3\height}{\strut Linking Similar Contributions (2)}& 
     
    \makebox[0pt][l]{\cellcolor{lightlightgrey}}%
    \raisebox{-0.5\height}{\strut%
    \begin{tikzpicture}
    \begin{axis}[testbar]
    \addplot[fill=cyan!40] coordinates{(50,0) };      % DL
    \addplot[fill=orange!40] coordinates{(50,0) }; % HB
    
    \end{axis}
    \end{tikzpicture}}
    
    &  
    \cellcolor{lightlightgrey}\raisebox{-0.4\height}{\strut \xmark} &  
    \cellcolor{lightlightgrey}\raisebox{-0.4\height}{\strut \xmark} & 
    \cellcolor{lightlightgrey}\raisebox{-0.4\height}{\strut 1/2} & 
    \cellcolor{lightlightgrey}\raisebox{-0.4\height}{\strut code change, PR} &
    
    \makebox[0pt][l]{\cellcolor{lightlightgrey}}%
    \raisebox{-0.5\height}{\strut%
    \begin{tikzpicture}
    \begin{axis}[testbar] 
    \addplot[fill=pink!40] coordinates{(50,0) }; % Java
    \addplot[fill=purple!40] coordinates{(50,0) };   % Language Independent
    \end{axis}
    \end{tikzpicture}}\\ 
     % ------------------------------------------------------------------------------------------------------------ end row
     
   \raisebox{-0.3\height}{\strut Predicting Salient-Class (1)}&
    
    \raisebox{-0.5\height}{\strut%
    \begin{tikzpicture}
    \begin{axis}[testbar] 
    \addplot[fill=red!40] coordinates{(100,0)}; % ML
    \end{axis}
    \end{tikzpicture}}
  
    &
    
    \raisebox{-0.4\height}{\strut \xmark} &
    \raisebox{-0.4\height}{\strut \xmark} &
    \raisebox{-0.4\height}{\strut 1/1} &
    \raisebox{-0.4\height}{\strut commit} &
    
    \raisebox{-0.5\height}{\strut%
    \begin{tikzpicture}
    \begin{axis}[testbar] 
    \addplot[fill=pink!40] coordinates{(100,0) }; % Java
    \end{axis}
    \end{tikzpicture}}\\
    \rulec
    % ------------------------------------------------------------------------------------------------------------ end row

     % CODE CHANGE CLASSIFICATION
     \multicolumn{7}{c}{\bf Code Change Classification} \\ \rulec
     
    \cellcolor{lightlightgrey}\raisebox{-0.3\height}{\strut Identifying Impactful Code Changes (2)}& 
     
    \makebox[0pt][l]{\cellcolor{lightlightgrey}}% 
    \raisebox{-0.5\height}{\strut%
    \begin{tikzpicture}
    \begin{axis}[testbar] 
    \addplot[fill=red!40] coordinates{(50,0)}; %ML
    \addplot[fill=orange!40] coordinates{(50,0) }; %HB
    \end{axis}
    \end{tikzpicture}}
    
     &  
     \cellcolor{lightlightgrey}\raisebox{-0.4\height}{\strut \xmark} & 
     \cellcolor{lightlightgrey}\raisebox{-0.4\height}{\strut \xmark} & 
     \cellcolor{lightlightgrey}\raisebox{-0.4\height}{\strut 1/2} & 
     \cellcolor{lightlightgrey}\raisebox{-0.4\height}{\strut commit} & 
     
    \makebox[0pt][l]{\cellcolor{lightlightgrey}}%
    \raisebox{-0.5\height}{\strut%
    \begin{tikzpicture}
    \begin{axis}[testbar] 
    \addplot[fill=pink!40] coordinates{(50,0) }; % Java
    \addplot[fill=yellow!40] coordinates{(50,0) }; % Multi
    \end{axis}
    \end{tikzpicture}}
     \\  
     % ------------------------------------------------------------------------------------------------------------ end row

         \raisebox{-0.3\height}{\strut Identifying Large-review-effort Code Changes (1)}& 
    
    \raisebox{-0.5\height}{\strut%
    \begin{tikzpicture}
    \begin{axis}[testbar] 
    \addplot[fill=red!40] coordinates{(100,0)}; %ML
    \end{axis}
    \end{tikzpicture}}
    
     &  
     \raisebox{-0.4\height}{\strut \xmark} &  
     \raisebox{-0.4\height}{\strut \xmark} & 
     \raisebox{-0.4\height}{\strut 1/1} & 
     \raisebox{-0.4\height}{\strut commit} & 
    
    \raisebox{-0.5\height}{\strut% 
    \begin{tikzpicture}
    \begin{axis}[testbar] 
    \addplot[fill=purple!40] coordinates{(100,0) }; % language independent
    \end{axis}
    \end{tikzpicture}}
     \\ 
     % ------------------------------------------------------------------------------------------------------------ end row
     
     \cellcolor{lightlightgrey}\raisebox{-0.3\height}{\strut Identifying Quickly Reviewable Changes (1)}& 

    \makebox[0pt][l]{\cellcolor{lightlightgrey}}%
    \raisebox{-0.5\height}{\strut%
    \begin{tikzpicture}
    \begin{axis}[testbar] 
    \addplot[fill=red!40] coordinates{(100,0)}; %ML
    \end{axis}
    \end{tikzpicture}}
    
     &  
     \cellcolor{lightlightgrey}\raisebox{-0.4\height}{\strut \xmark} &  
     \cellcolor{lightlightgrey}\raisebox{-0.4\height}{\strut \xmark} & 
     \cellcolor{lightlightgrey}\raisebox{-0.4\height}{\strut 1/1} & 
     \cellcolor{lightlightgrey}\raisebox{-0.4\height}{\strut PR} & 
     
    \makebox[0pt][l]{\cellcolor{lightlightgrey}}% 
    \raisebox{-0.5\height}{\strut%
    \begin{tikzpicture}
    \begin{axis}[testbar]
    \addplot[fill=pink!40] coordinates{(100,0) }; % Java
    \end{axis}
    \end{tikzpicture}}
     \\ 
     % ------------------------------------------------------------------------------------------------------------ end row

      Predicting Code Changes Approval, Merge, or Need for review (11)&
      
    \raisebox{-0.3\height}{\strut% 
    \begin{tikzpicture}
    \begin{axis}[testbar] 
    \addplot[fill=cyan!40] coordinates{(54.5,0) };    % DL     6/11 54.5%
    \addplot[fill=red!40] coordinates{(27.5,0)};       % ML     3/11 27.5%
    \addplot[fill=gray!40] coordinates{(18,0) };       % Other 2/11 (search-based) 18%
    \end{axis}
    \end{tikzpicture}}
    
     & 
     \raisebox{-0.1\height}{\strut 3/11} & 
     \raisebox{-0.1\height}{\strut 4/11} & 
     \raisebox{-0.1\height}{\strut 11/11} & 
     \raisebox{-0.1\height}{\strut diff hunk, method, PR } & 
     
    \raisebox{-0.3\height}{\strut%
    \begin{tikzpicture}
    \begin{axis}[testbar] 
    \addplot[fill=pink!40] coordinates{(36.5,0) }; % Java 4/11 36.5%
    \addplot[fill=yellow!40] coordinates{(27,0) }; % Multi 3/11 27%
    \addplot[fill=purple!40] coordinates{(37.5,0) }; % language independent 4/11 36.5%
    \end{axis}
    \end{tikzpicture}}\\ 
     \rulec
     % ------------------------------------------------------------------------------------------------------------ end row

       % CODE CHANGE QUALITY CHECK
     \multicolumn{7}{c}{\bf Code Change Quality Check} \\ \rulec
     % ------------------------------------------------------------------------------------------------------------ end row
     
     \cellcolor{lightlightgrey}\raisebox{-0.3\height}{\strut Checking Design Patterns Consistency (1)}&
    
    \makebox[0pt][l]{\cellcolor{lightlightgrey}}%
    \raisebox{-0.5\height}{\strut% 
    \begin{tikzpicture}
    \begin{axis}[testbar] 
    \addplot[fill=orange!40] coordinates{(100,0) }; %HB
    \end{axis}
    \end{tikzpicture}}
    
     & 
     \cellcolor{lightlightgrey}\raisebox{-0.4\height}{\strut \xmark} & 
     \cellcolor{lightlightgrey}\raisebox{-0.4\height}{\strut \xmark} & 
     \cellcolor{lightlightgrey}\raisebox{-0.4\height}{\strut \xmark} & 
     \cellcolor{lightlightgrey}\raisebox{-0.4\height}{\strut file} & 
     
    \makebox[0pt][l]{\cellcolor{lightlightgrey}}% 
    \raisebox{-0.5\height}{\strut%
    \begin{tikzpicture}
    \begin{axis}[testbar] 
    \addplot[fill=pink!40] coordinates{(100,0) }; % Java
    \end{axis}
    \end{tikzpicture}}
     \\ 
     % ------------------------------------------------------------------------------------------------------------ end row
     
    \raisebox{-0.3\height}{\strut Generating Review Comments (11)}& 
     
    \raisebox{-0.5\height}{\strut%
    \begin{tikzpicture}
    \begin{axis}[testbar] 
    \addplot[fill=cyan!40] coordinates{(91,0) };       % DL 10/11 91%
    \addplot[fill=violet!40] coordinates{(9,0) };     % IR  1/11 9%
    \end{axis}
    \end{tikzpicture}}
    
     & 
     \raisebox{-0.4\height}{\strut 9/11} & 
     \raisebox{-0.4\height}{\strut 10/11} & 
     \raisebox{-0.4\height}{\strut 10/11} & 
     \raisebox{-0.4\height}{\strut code change, diff hunk, method } & 
     
    \raisebox{-0.5\height}{\strut%
    \begin{tikzpicture}
    \begin{axis}[testbar] 
    \addplot[fill=pink!40] coordinates{(36.5,0) }; % Java    4/11 36.5
    \addplot[fill=yellow!40] coordinates{(63.5,0) }; % Multi 7/11 63.5%
    \end{axis}
    \end{tikzpicture}}
     \\  
     % ------------------------------------------------------------------------------------------------------------ end row

    \cellcolor{lightlightgrey}\raisebox{-0.3\height}{\strut Identifying Clone Refactoring Opportunities (1)}&
    
    \makebox[0pt][l]{\cellcolor{lightlightgrey}}%
    \raisebox{-0.5\height}{\strut%
    \begin{tikzpicture}
    \begin{axis}[testbar] 
    \addplot[fill=orange!40] coordinates{(100,0) }; %HB
    \end{axis}
    \end{tikzpicture}}
    
     & 
     \cellcolor{lightlightgrey}\raisebox{-0.4\height}{\strut \xmark} & 
     \cellcolor{lightlightgrey}\raisebox{-0.4\height}{\strut \xmark} & 
     \cellcolor{lightlightgrey}\raisebox{-0.4\height}{\strut \xmark} & 
     \cellcolor{lightlightgrey}\raisebox{-0.4\height}{\strut PR} & 
     
    \makebox[0pt][l]{\cellcolor{lightlightgrey}}%
    \raisebox{-0.5\height}{\strut%
    \begin{tikzpicture}
    \begin{axis}[testbar] 
    \addplot[fill=pink!40] coordinates{(100,0) }; % Java
    \end{axis}
    \end{tikzpicture}}
     \\  
     % ------------------------------------------------------------------------------------------------------------ end row
     
    \raisebox{-0.3\height}{\strut Predicting Code Defectiveness (2)}&
      
    \raisebox{-0.5\height}{\strut%
    \begin{tikzpicture}
    \begin{axis}[testbar] 
    \addplot[fill=red!40] coordinates{(50,0)}; %ML
    \addplot[fill=orange!40] coordinates{(50,0) }; %HB
    \end{axis}
    \end{tikzpicture}}
    
     & 
     \raisebox{-0.4\height}{\strut \xmark} & 
     \raisebox{-0.4\height}{\strut \xmark} & 
     \raisebox{-0.4\height}{\strut 1/2} & 
     \raisebox{-0.4\height}{\strut file, PR} & 
    
    \raisebox{-0.5\height}{\strut% 
    \begin{tikzpicture}
    \begin{axis}[testbar] 
    \addplot[fill=purple!40] coordinates{(100,0) }; % language independent
    \end{axis}
    \end{tikzpicture}}
     \\  
    % ------------------------------------------------------------------------------------------------------------ end row

    \cellcolor{lightlightgrey}\raisebox{-0.3\height}{\strut Predicting Problematic Code Elements (3)}& 
    
    \makebox[0pt][l]{\cellcolor{lightlightgrey}}%
    \raisebox{-0.5\height}{\strut%
    \begin{tikzpicture}
    \begin{axis}[testbar] 
    \addplot[fill=cyan!40] coordinates{(66.5,0) };   % DL
    \addplot[fill=red!40] coordinates{(33.5,0)};      % ML
    \end{axis}
    \end{tikzpicture}}
    
     & 
     \cellcolor{lightlightgrey}\raisebox{-0.4\height}{\strut 2/3} & 
     \cellcolor{lightlightgrey}\raisebox{-0.4\height}{\strut 2/3} & 
     \cellcolor{lightlightgrey}\raisebox{-0.4\height}{\strut 3/3} & 
     \cellcolor{lightlightgrey}\raisebox{-0.4\height}{\strut code line, file, PR} & 
     
    \makebox[0pt][l]{\cellcolor{lightlightgrey}}%
    \raisebox{-0.5\height}{\strut%
    \begin{tikzpicture}
    \begin{axis}[testbar] 
    \addplot[fill=pink!40] coordinates{(33.5,0) };     % Java 1/3
    \addplot[fill=yellow!40] coordinates{(66.5,0) }; % Multi 2/3
    \end{axis}
    \end{tikzpicture}}
     \\  
     % ------------------------------------------------------------------------------------------------------------ end row
      
    \raisebox{-0.3\height}{\strut Reviewing Code Formatting Violations (1)}&
     
    \raisebox{-0.5\height}{\strut% 
    \begin{tikzpicture}
    \begin{axis}[testbar] 
    \addplot[fill=red!40] coordinates{(100,0)}; %ML
    \end{axis}
    \end{tikzpicture}}
    
     & 
     \raisebox{-0.4\height}{\strut \xmark} & 
     \raisebox{-0.4\height}{\strut \xmark} & 
     \raisebox{-0.4\height}{\strut 1/1} & 
     \raisebox{-0.4\height}{\strut file} & 
    
    \raisebox{-0.5\height}{\strut%
    \begin{tikzpicture}
    \begin{axis}[testbar] 
    \addplot[fill=gray!40] coordinates{(100,0) }; % Other JavaScript
    \end{axis}
    \end{tikzpicture}}
     \\ 
     % ------------------------------------------------------------------------------------------------------------ end row
     
   \cellcolor{lightlightgrey}\raisebox{-0.3\height}{\strut Reviewing via Static Analysis (1)}& 
     
    \makebox[0pt][l]{\cellcolor{lightlightgrey}}% 
    \raisebox{-0.5\height}{\strut%
    \begin{tikzpicture}
    \begin{axis}[testbar] 
    \addplot[fill=orange!40] coordinates{(100,0) }; %HB
    \end{axis}
    \end{tikzpicture}}
    
     & 
     \cellcolor{lightlightgrey}\raisebox{-0.4\height}{\strut \xmark} & 
     \cellcolor{lightlightgrey}\raisebox{-0.4\height}{\strut \xmark} & 
     \cellcolor{lightlightgrey}\raisebox{-0.4\height}{\strut \xmark} & 
     \cellcolor{lightlightgrey}\raisebox{-0.4\height}{\strut PR} & 
    
    \makebox[0pt][l]{\cellcolor{lightlightgrey}}% 
    \raisebox{-0.5\height}{\strut%
    \begin{tikzpicture}
    \begin{axis}[testbar] 
    \addplot[fill=pink!40] coordinates{(100,0) }; % Java
    \end{axis}
    \end{tikzpicture}}
     \\
     \rulec
     % ------------------------------------------------------------------------------------------------------------ end row

   % CODE REVIEW SENTIMENT ANALYIS
   \multicolumn{7}{c}{\bf Code Review Sentiment Analysis} \\ \rulec
    % ------------------------------------------------------------------------------------------------------------ end row
     
    \raisebox{-0.3\height}{\strut Classifying the Sentiment of Review Comments (1)}& 
      
    \raisebox{-0.5\height}{\strut%
    \begin{tikzpicture}
    \begin{axis}[testbar] 
    \addplot[fill=red!40] coordinates{(100,0)}; %ML
    \end{axis}
    \end{tikzpicture}}
    
     & 
     \raisebox{-0.4\height}{\strut \xmark} & 
     \raisebox{-0.4\height}{\strut \xmark} & 
     \raisebox{-0.4\height}{\strut 1/1} & 
     \raisebox{-0.4\height}{\strut review comment} & 
     
     \raisebox{-0.5\height}{\strut%
    \begin{tikzpicture}
    \begin{axis}[testbar] 
    \addplot[fill=purple!40] coordinates{(100,0) }; % language independent
    \end{axis}
    \end{tikzpicture}}
     \\ 
     % ------------------------------------------------------------------------------------------------------------ end row
     
    \cellcolor{lightlightgrey}\raisebox{-0.3\height}{\strut Identifying “Pushback” Feelings in Reviews (1)}& 
    
    \makebox[0pt][l]{\cellcolor{lightlightgrey}}% 
    \raisebox{-0.5\height}{\strut%
    \begin{tikzpicture}
    \begin{axis}[testbar] 
    \addplot[fill=orange!40] coordinates{(100,0)}; % HB
    \end{axis}
    \end{tikzpicture}}
    
     & 
     \cellcolor{lightlightgrey}\raisebox{-0.4\height}{\strut \xmark} & 
     \cellcolor{lightlightgrey}\raisebox{-0.4\height}{\strut \xmark} & 
     \cellcolor{lightlightgrey}\raisebox{-0.4\height}{\strut \xmark} & 
     \cellcolor{lightlightgrey}\raisebox{-0.4\height}{\strut code review} & 
    
    \makebox[0pt][l]{\cellcolor{lightlightgrey}}% 
    \raisebox{-0.5\height}{\strut%
    \begin{tikzpicture}
    \begin{axis}[testbar] 
    \addplot[fill=purple!40] coordinates{(100,0) }; % language independent
    \end{axis}
    \end{tikzpicture}}
     \\ 
     % ------------------------------------------------------------------------------------------------------------ end row
     
    \raisebox{-0.75\height}{\strut Identifying Toxic/Uncivil Code Review Comments (4)}& 
     
    \raisebox{-0.85\height}{\strut%
    \begin{tikzpicture}
    \begin{axis}[testbar] 
    \addplot[fill=cyan!40] coordinates{(100,0)}; % DL
    \end{axis}
    \end{tikzpicture}}
    
     & 
     \raisebox{-0.75\height}{\strut 4/4} & 
     \raisebox{-0.75\height}{\strut \xmark} & 
     \raisebox{-0.75\height}{\strut 4/4} & 
     email, review comment, sentence & 
    
    \raisebox{-0.85\height}{\strut% 
    \begin{tikzpicture}
    \begin{axis}[testbar] 
    \addplot[fill=purple!40] coordinates{(100,0) }; % language independent
    \end{axis}
    \end{tikzpicture}}
     \\  
     % ------------------------------------------------------------------------------------------------------------ end row
     
    \cellcolor{lightlightgrey}\raisebox{-0.3\height}{\strut Rephrasing Toxic/Uncivil Comments (1)}& 
    
    \makebox[0pt][l]{\cellcolor{lightlightgrey}}%
    \raisebox{-0.5\height}{\strut%
    \begin{tikzpicture}
    \begin{axis}[testbar] 
    \addplot[fill=cyan!40] coordinates{(100,0) }; % DL
    \end{axis}
    \end{tikzpicture}}
    
     & 
     \cellcolor{lightlightgrey}\raisebox{-0.4\height}{\strut 1/1} & 
     \cellcolor{lightlightgrey}\raisebox{-0.4\height}{\strut 1/1} & 
     \cellcolor{lightlightgrey}\raisebox{-0.4\height}{\strut 1/1} & 
     \cellcolor{lightlightgrey}\raisebox{-0.4\height}{\strut review comment} & 
    
    \makebox[0pt][l]{\cellcolor{lightlightgrey}}%
    \raisebox{-0.5\height}{\strut% 
    \begin{tikzpicture}
    \begin{axis}[testbar] 
    \addplot[fill=purple!40] coordinates{(100,0) }; % language independent
    \end{axis}
    \end{tikzpicture}}
     \\ 
     \rulec 
     % ------------------------------------------------------------------------------------------------------------ end row
     
    % RETRIVIAL OF SIMILAR CR/CC
   \multicolumn{7}{c}{\bf Retrieval of Similar CR/CC} \\ \rulec
   % ------------------------------------------------------------------------------------------------------------ end row
     
    \raisebox{-1.5\height}{\strut Augmenting Reviews (8)}& 
      
    \raisebox{-1.45\height}{\strut%
    \begin{tikzpicture}
    \begin{axis}[testbar] 
    \addplot[fill=cyan!40] coordinates{(37.5,0) };       % DL 3/8 37.5%
    \addplot[fill=red!40] coordinates{(25,0)};          % ML 2/8 25
    \addplot[fill=violet!40] coordinates{(25,0) };     % IR 2/8 25
    \addplot[fill=gray!40] coordinates{(12.5,0) };       % Other search-based 1/8  12.5
    \end{axis}
    \end{tikzpicture}}
    
     & 
     \raisebox{-1.5\height}{\strut \xmark} & 
     \raisebox{-1.5\height}{\strut \xmark} & 
     \raisebox{-1.5\height}{\strut 4/8} & 
     code change, code review, code snippet, diff hunk, review comment& 
     
    \raisebox{-1.45\height}{\strut% 
    \begin{tikzpicture}
    \begin{axis}[testbar] 
    \addplot[fill=pink!40] coordinates{(50,0) }; % Java 4/8 37.5%
    \addplot[fill=yellow!40] coordinates{(12.5,0) };   % Multiple Languages 1/8
    \addplot[fill=purple!40] coordinates{(25,0) }; % language independent 2/8
    \addplot[fill=gray!40] coordinates{(12.5,0) }; % Other C 1/8
    \end{axis}
    \end{tikzpicture}}
     \\ 
    % ------------------------------------------------------------------------------------------------------------ end row
     
    \cellcolor{lightlightgrey}\raisebox{-0.3\height}{\strut Mining Code Improvement Patterns (1)}& 
    
    \makebox[0pt][l]{\cellcolor{lightlightgrey}}% 
    \raisebox{-0.5\height}{\strut%
    \begin{tikzpicture}
    \begin{axis}[testbar] 
    \addplot[fill=gray!40] coordinates{(100,0) }; % Other
    \end{axis}
    \end{tikzpicture}}
    
     & 
     \cellcolor{lightlightgrey}\raisebox{-0.4\height}{\strut \xmark} & 
     \cellcolor{lightlightgrey}\raisebox{-0.4\height}{\strut \xmark} & 
     \cellcolor{lightlightgrey}\raisebox{-0.4\height}{\strut 1/1} & 
     \cellcolor{lightlightgrey}\raisebox{-0.4\height}{\strut diff hunk} & 
    
    \makebox[0pt][l]{\cellcolor{lightlightgrey}}%
    \raisebox{-0.5\height}{\strut% 
    \begin{tikzpicture}
    \begin{axis}[testbar] 
    \addplot[fill=gray!40] coordinates{(100,0) }; %other Python
    \end{axis}
    \end{tikzpicture}}
     \\ 
     \rulec 
     % ------------------------------------------------------------------------------------------------------------ end row
     
     % REVISED CODE GENERATION
   \multicolumn{7}{c}{\bf Revised Code Generation} \\ \rulec
   % ------------------------------------------------------------------------------------------------------------ end row
   
     Implementing the Code Change Requested by a Reviewer (11)& 
    
    \raisebox{-0.3\height}{\strut%
    \begin{tikzpicture}
    \begin{axis}[testbar] 
    \addplot[fill=cyan!40] coordinates{(100,0) }; %DL
    \end{axis}
    \end{tikzpicture}}
    
     & 
     \raisebox{-0.3\height}{\strut 8/11} & 
     \raisebox{-0.3\height}{\strut 9/11} & 
     \raisebox{-0.3\height}{\strut 11/11} & 
    code change, diff hunk, method & 
   
    \raisebox{-0.3\height}{\strut% 
    \begin{tikzpicture}
    \begin{axis}[testbar] 
    \addplot[fill=pink!40] coordinates{(36.5,0) }; % Java 4/11 36.5
    \addplot[fill=yellow!40] coordinates{(63.5,0) }; % Multi 7/11 63.5
    \end{axis}
    \end{tikzpicture}}
     \\ 
     % ------------------------------------------------------------------------------------------------------------ end row
     
    \cellcolor{lightlightgrey}\raisebox{-0.3\height}{\strut Predicting the Code Output of the Review Process (4)}& 
    
    \makebox[0pt][l]{\cellcolor{lightlightgrey}}%  
    \raisebox{-0.5\height}{\strut%
    \begin{tikzpicture}
    \begin{axis}[testbar] 
    \addplot[fill=cyan!40] coordinates{(100,0) }; %DL
    \end{axis}
    \end{tikzpicture}}
    
     & 
     \cellcolor{lightlightgrey}\raisebox{-0.4\height}{\strut 2/3} & 
     \cellcolor{lightlightgrey}\raisebox{-0.4\height}{\strut 2/3} & 
     \cellcolor{lightlightgrey}\raisebox{-0.4\height}{\strut 4} & 
     \cellcolor{lightlightgrey}\raisebox{-0.4\height}{\strut method} & 
    
    \makebox[0pt][l]{\cellcolor{lightlightgrey}}%
    \raisebox{-0.5\height}{\strut% 
    \begin{tikzpicture}
    \begin{axis}[testbar] 
    \addplot[fill=pink!40] coordinates{(100,0) }; % Java
    \end{axis}
    \end{tikzpicture}}
     \\ 
     \rulec
     % ------------------------------------------------------------------------------------------------------------ end row

   % TIME MANAGEMENT
   \multicolumn{7}{c}{\bf Time Management} \\ \rulec
   % ------------------------------------------------------------------------------------------------------------ end row
   
    \raisebox{-0.3\height}{\strut Identifying Blocking Actors in Pull Requests (1)}&
     
    \raisebox{-0.5\height}{\strut% 
    \begin{tikzpicture}
    \begin{axis}[testbar] 
    \addplot[fill=orange!40] coordinates{(100,0) }; %HB
    \end{axis}
    \end{tikzpicture}}
    
     & 
     \raisebox{-0.4\height}{\strut \xmark} & 
     \raisebox{-0.4\height}{\strut \xmark} & 
     \raisebox{-0.4\height}{\strut \xmark} & 
     \raisebox{-0.4\height}{\strut PR} & 
     
    \raisebox{-0.5\height}{\strut%
    \begin{tikzpicture}
    \begin{axis}[testbar] 
    \addplot[fill=purple!40] coordinates{(100,0) }; % language independent
    \end{axis}
    \end{tikzpicture}}
     \\
    % ------------------------------------------------------------------------------------------------------------ end row
   
   \cellcolor{lightlightgrey}Predicting Pull Request/Code Review Completion Time (5)&
    
    \makebox[0pt][l]{\cellcolor{lightlightgrey}}%
    \raisebox{-0.3\height}{\strut%
    \begin{tikzpicture}
    \begin{axis}[testbar] 
    \addplot[fill=red!40] coordinates{(80,0)};    %ML  4/5
    \addplot[fill=orange!40] coordinates{(20,0) };   % HB 1/5
    \end{axis}
    \end{tikzpicture}}
    
     & 
     \cellcolor{lightlightgrey}\raisebox{-0.2\height}{\strut \xmark} & 
     \cellcolor{lightlightgrey}\raisebox{-0.2\height}{\strut \xmark} & 
     \cellcolor{lightlightgrey}\raisebox{-0.2\height}{\strut 4/5} & 
     \cellcolor{lightlightgrey}\raisebox{-0.2\height}{\strut code change, commit, PR } & 
    
    \makebox[0pt][l]{\cellcolor{lightlightgrey}}%
    \raisebox{-0.3\height}{\strut% 
    \begin{tikzpicture}
    \begin{axis}[testbar] 
    \addplot[fill=purple!40] coordinates{(100,0) }; % language independent
    \end{axis}
    \end{tikzpicture}}
     \\
     % ------------------------------------------------------------------------------------------------------------ end row

    \raisebox{-0.3\height}{\strut Prioritizing Review Requests (3)}&
    
    \raisebox{-0.5\height}{\strut%
    \begin{tikzpicture}
    \begin{axis}[testbar] 
    \addplot[fill=cyan!40] coordinates{(33.33,0) };  % DL
    \addplot[fill=red!40] coordinates{(33.33,0)};   % ML
    \addplot[fill=gray!40] coordinates{(33.33,0) };  % Other 1/3
    \end{axis}
    \end{tikzpicture}}
    
     & 
     \raisebox{-0.4\height}{\strut \xmark} & 
     \raisebox{-0.4\height}{\strut \xmark} & 
     \raisebox{-0.4\height}{\strut 2/3} & 
     \raisebox{-0.4\height}{\strut code change, PR} & 
    
    \raisebox{-0.5\height}{\strut% 
    \begin{tikzpicture}
    \begin{axis}[testbar] 
    \addplot[fill=purple!40] coordinates{(100,0) }; % language independent
    \end{axis}
    \end{tikzpicture}}
     \\ 
     \rulec 
    % ------------------------------------------------------------------------------------------------------------ end row
     
      % OTHER
   \multicolumn{7}{c}{\bf Other} \\ \rulec
   % ------------------------------------------------------------------------------------------------------------ end row
   
   \cellcolor{lightlightgrey}Classifying the Goal of a Review Comment or the Type of Change Triggered by a Comment (3)&
   
    \makebox[0pt][l]{\cellcolor{lightlightgrey}}%
    \raisebox{-0.3\height}{\strut% 
    \begin{tikzpicture}
    \begin{axis}[testbar] 
    \addplot[fill=cyan!40] coordinates{(33.5,0)}; %DL
    \addplot[fill=red!40] coordinates{(66.5,0)};   %ML
    \end{axis}
    \end{tikzpicture}}
    
     & 
     \cellcolor{lightlightgrey}\raisebox{-0.2\height}{\strut 1/3} & 
     \cellcolor{lightlightgrey}\raisebox{-0.2\height}{\strut 1/3} & 
     \cellcolor{lightlightgrey}\raisebox{-0.2\height}{\strut 3/3} & 
     \cellcolor{lightlightgrey}\raisebox{-0.2\height}{\strut code change, review comment} & 
    
    \makebox[0pt][l]{\cellcolor{lightlightgrey}}% 
    \raisebox{-0.3\height}{\strut%
    \begin{tikzpicture}
    \begin{axis}[testbar] 
   \addplot[fill=pink!40] coordinates{(33.3,0) };       % Java
    \addplot[fill=purple!40] coordinates{(33.3,0) };  % language independent
    \addplot[fill=gray!40] coordinates{(33.4,0) };     % other (python)
    \end{axis}
    \end{tikzpicture}}
     \\
     % ------------------------------------------------------------------------------------------------------------ end row
     
     \raisebox{-0.3\height}{\strut Configuring Static Code Analysis Tools (1)}&
     
    \raisebox{-0.5\height}{\strut%
    \begin{tikzpicture}
    \begin{axis}[testbar] 
    \addplot[fill=red!40] coordinates{(100,0)}; %ML
    \end{axis}
    \end{tikzpicture}}
    
     & 
     \raisebox{-0.4\height}{\strut \xmark} & 
     \raisebox{-0.4\height}{\strut \xmark} & 
     \raisebox{-0.4\height}{\strut 1/1} & 
     \raisebox{-0.4\height}{\strut review comment} & 
 
    \raisebox{-0.5\height}{\strut%    
    \begin{tikzpicture}
    \begin{axis}[testbar] 
    \addplot[fill=pink!40] coordinates{(100,0) }; % Java
    \end{axis}
    \end{tikzpicture}}
     \\
   % ------------------------------------------------------------------------------------------------------------ end row
   
    \cellcolor{lightlightgrey}\raisebox{-0.3\height}{\strut Partitioning Static Analysis Warnings (1)}&
    
    \makebox[0pt][l]{\cellcolor{lightlightgrey}}%
    \raisebox{-0.5\height}{\strut% 
    \begin{tikzpicture}
    \begin{axis}[testbar] 
    \addplot[fill=gray!40] coordinates{(100,0)}; % Other
    \end{axis}
    \end{tikzpicture}}
    
     & 
     \cellcolor{lightlightgrey}\raisebox{-0.4\height}{\strut \xmark} & 
     \cellcolor{lightlightgrey}\raisebox{-0.4\height}{\strut \xmark} & 
     \cellcolor{lightlightgrey}\raisebox{-0.4\height}{\strut \xmark} & 
     \cellcolor{lightlightgrey}\raisebox{-0.4\height}{\strut code snippet} & 
    
    \makebox[0pt][l]{\cellcolor{lightlightgrey}}%
    \raisebox{-0.5\height}{\strut%    
    \begin{tikzpicture}
    \begin{axis}[testbar] 
    \addplot[fill=gray!40] coordinates{(100,0) }; % Other C
    \end{axis}
    \end{tikzpicture}}
     \\
    % ------------------------------------------------------------------------------------------------------------ end row
        
    \raisebox{-0.3\height}{\strut Recommending Reviewers (36)}&
     
    \raisebox{-0.5\height}{\strut%
    \begin{tikzpicture}
    \begin{axis}[testbar] 
   \addplot[fill=cyan!40] coordinates{(5.5,0) };          % DL 2 5.5%
   \addplot[fill=red!40] coordinates{(28,0)};      % ML 10 27.7%
   \addplot[fill=orange!40] coordinates{(36,0) }; % HB 13 36.1%
   \addplot[fill=gray!40] coordinates{(30.5,0) };       % Other 5 GB + 1 new algorithm + 5 SB = 11 -> 30.5%
    \end{axis}
    \end{tikzpicture}}
    
     & 
     \raisebox{-0.4\height}{\strut 1/36} & 
     \raisebox{-0.4\height}{\strut \xmark} & 
     \raisebox{-0.4\height}{\strut 23/36} & 
     \raisebox{-0.4\height}{\strut commit, patch, PR} & 
   
    \raisebox{-0.85\height}{\strut% 
    \begin{tikzpicture}
    \begin{axis}[testbar] 
    \addplot[fill=pink!40] coordinates{(8.5,0) }; % Java 3/36 8.33%
    \addplot[fill=yellow!40] coordinates{(5.5,0) }; % Multi 2/36 5.5%
    \addplot[fill=purple!40] coordinates{(86,0) }; % language independent  31/36  86.1
    \end{axis}
    \end{tikzpicture}}
     \\
    % ------------------------------------------------------------------------------------------------------------ end row
     
    \cellcolor{lightlightgrey}\raisebox{-0.3\height}{\strut Visualizing Code Changes (4)}&
    
    \makebox[0pt][l]{\cellcolor{lightlightgrey}}% 
    \raisebox{-0.5\height}{\strut%
    \begin{tikzpicture}
    \begin{axis}[testbar] 
    \addplot[fill=orange!40] coordinates{(50,0) }; %HB
    \addplot[fill=gray!40] coordinates{(50,0) }; % O
    \end{axis}
    \end{tikzpicture}}
    
     & 
     \cellcolor{lightlightgrey}\raisebox{-0.4\height}{\strut \xmark} & 
     \cellcolor{lightlightgrey}\raisebox{-0.4\height}{\strut \xmark} & 
     \cellcolor{lightlightgrey}\raisebox{-0.4\height}{\strut \xmark} & 
     \cellcolor{lightlightgrey}\raisebox{-0.4\height}{\strut commit, diff hunk, PR} & 
     
    \makebox[0pt][l]{\cellcolor{lightlightgrey}}% 
    \raisebox{-0.5\height}{\strut%
    \begin{tikzpicture}
    \begin{axis}[testbar] 
    \addplot[fill=pink!40] coordinates{(50,0) }; % Java
    \addplot[fill=yellow!40] coordinates{(50,0) }; % Multi
    \end{axis}
    \end{tikzpicture}}
     \\  
    % ------------------------------------------------------------------------------------------------------------ end row

    \rulec
    \end{tabular}
    }
\end{table}

\subsection{RQ$_2$: What are the under-the-hood solutions behind the techniques and tools proposed for code review automation?}

\tabref{tab:rq2} summarizes the under-the-hood solutions behind the techniques proposed in the literature for code review automation. The ``Task'' column reports the list of automated tasks, with the number in parenthesis representing the number of papers (out of the considered \selected) presenting an automation solution for such a task. For each task $T_i$, in the ``Approach'' column the bar chart depicts the percentage of DL-based, ML-based, IR-based, and Heuristic-based techniques out of those automating $T_i$. Approaches not relying on any of these four techniques are grouped into the ``Other'' categories (\eg data-flow analysis \cite{tukaram:scam2013} or visualization techniques \cite{menarini:ase2017}). With heuristic-based techniques we refer to hand-crafted techniques which are usually composed by multiple steps (\eg building a traceability graph and defining a specific metric to identify the best-suited reviewer for a given code change \cite{sulun:icpmdase2019}). 

For approaches based on DL/ML, the ``Training'' column shows whether they underwent (i) a pre-training on natural language corpus (``PT NL''); (ii) a pre-training on a code corpus (``PT code''); and (iii) a fine-tuning (``FT''). While the pre-training procedures are typical of DL-based techniques, with fine-tuning we also indicate the standard training of classic ML algorithms (\eg training a classifier to identify design-impactful changes on a labeled dataset \cite{uchoa:msr2021}). For each of these three training procedures, a \xmark indicates that none of the corresponding papers adopts it, otherwise a fraction is used to report the number of papers employing it. 

The ``Granularity'' column indicates, for a given task, the type of ``entities'' for which automation solutions have been proposed. For example, among the 11 techniques aimed at commenting on source code by posting natural language comments as a human would do (\ie \emph{generating review comments} task), some of them work at \emph{code change} granularity (\ie they take as input the whole code diff of a PR), others consider a specific \emph{diff hunk} (\ie only a specific part of the change, possibly spanning multiple functions), and the remaining ones work on a single \emph{function} impacted by the change (\ie they comment on one changed function).

Finally, for each task, the ``Language'' column depicts, using again a bar chart, the percentage of proposed automation techniques providing support for a specific programming language. Since Java was by far the most popular language, a specific color has been assigned to it (see \tabref{tab:rq2}'s caption), while other colors are used to indicate techniques (i) supporting multiple languages, (ii) being language-independent, (iii) or being specific for a single language which is not Java. When we report an approach as only supporting a specific or multiple languages, this does not mean that the approach cannot be adapted to other languages. This is something we did not assess, since it would require a deep understanding of all technicalities behind each approach, something which is not always easy to grasp from the paper's reading. For example, a DL model trained and tested on Java code to support a specific task, is labeled as ``Java only'' despite, with a reasonable effort, the approach could probably be trained on another languages keeping similar performance. Basically, we considered the languages on which the approaches can be used out of the box.

\subsubsection{Approach} There is the clear distinction between the underlying solutions adopted by techniques automating classification \emph{vs} generative tasks. For the former (\eg \emph{classifying the usefulness of review comments}, \emph{predicting salient-class}, \emph{identifying impactful code changes}), ML-based solutions (red bars in \tabref{tab:rq2}) are the most popular ones (36\%), followed by heuristic-based (24\%) and DL-based (21\%) techniques. Other solutions account for the remaining 19\% of techniques automating classification tasks, with none of them relying on IR. The situation is quite different for generative tasks from which the generation of textual output is expected (\eg \emph{generating review comments}, \emph{implementing the code change requested by a reviewer}, \emph{predicting the code output of the review process}). In this case, DL-based solutions are by far the most employed (78\%), followed by IR-based ones (11\%) which can identify relevant content in a knowledge base and use it as output. For example, in the \emph{generating review comments} task, the approach can take a piece of code to review $C_i$, find in a knowledge base the code $C_j$ being most similar to $C_i$, and reuse the reviewers' comments posted for $C_j$ when reviewing $C_i$.

For other types of tasks which cannot really be categorized as classification or generative tasks (\eg \emph{checking design patterns consistency}, \emph{reviewing via static analysis}, \emph{visualizing code changes}), there is no clear trend which can be observed, with all type of solutions being explored.

Interestingly is also to comment on the strongly increasing adoption of DL-based techniques for code review automation. If we focus on the last five years considered in our SLR (2020 to 2024), we find that in 2020 and in 2021 DL models have been exploited in 20\% (2/10) and in 17\% (2/12), respectively, of the papers in our SLR. From 2022, instead, we observe a strong increase in the adoption of DL-based solutions, with 42\% in 2022 (11/26), 68\% in 2023 (13/19), and 63\% in 2024 (17/27).

\subsubsection{Training procedures} Out of the 46 DL-based solution, 34 use some form of pre-training. The idea of pre-training is mostly to teach the DL model the language of interest, by performing a task-agnostic training. For example, a model meant to automatically generate review comments may be pre-trained on a corpus of natural language and code instances via the Masked Language Modeling (MLM) pre-training objective, providing the model with a sentence as input (\eg an English sentence or a Java statement) having 15\% of its tokens masked, with the model in charge of guessing the masked tokens. Of the 34 automated techniques using pre-training, 24 start from an already pre-trained model (\eg Code Llama \cite{roziere2023code}, CodeT5 \cite{wang-etal-2021-codet5}, RoBERTa \cite{liu:arxiv2019}), while the remaining ones pre-train their own model. In both cases, the pre-training usually involves both natural language and code (\ie bi-modal pre-training): This is visible in \tabref{tab:rq2} by comparing the number of papers using a model pre-trained on natural-language (column ``PT NL'') with those exploiting a model pre-trained on code (column ``PT code''). For example, out of the 11 papers \emph{predicting code changes approval, merge, or need for review}, 4 use a pre-trained model, 3 of which pre-trained on bi-modal data (+1 only code). Similarly, when looking at the ones \emph{implementing the code change requested by a reviewer}, 9 of the 11 papers use a pre-trained model, in 8 cases pre-trained on bi-modal data. Such a choice may have been driven by the empirical evidence showing that pre-training on natural language helps for code-related tasks as well \cite{tufano:ast2022}. There is only one exception to this trend: All four DL-based techniques aimed at \emph{identifying toxic/uncivil code review comments} exploit pre-training only on natural language. This is a sensible choice considering that the tackled task does not foresee the model manipulating code elements.

It is also worth mentioning the radical changes observed in the usage of pre-training in recent years. First, before 2022, we found no work on code review automation using a pre-trained DL model. Second, in 2022, out of the eight automated solutions exploiting pre-training, only one (12.5\%) used an already pre-trained models (in the other cases, the authors of the technique pre-trained their own model). In 2023 and 2024 this trend radically changed, with only 2 of the 26 proposed techniques relying on a pre-trained model \cite{vijayvergiya:aiware2024,froemmgen:icse2024} exploiting a model pre-trained by the authors themself. Such a trend is easily explained by the always-increasing availability of open source pre-trained models in websites such as HuggingFace \cite{wolf:arxiv2019}.

Concerning the fine-tuning (\ie the training of a model aimed at specializing it to the target task), all but three ML/DL-based techniques exploit it. The first two are those \emph{assessing review quality through biometrics}, which use already trained models to interpret in real time the biometric information collected by dedicated devices (\eg heart rate variability and pupillary response are captured and interpreted as ``mental workload''). The third one is the work by Widyasari \etal \cite{widyasari:tosem2024} exploiting prompt engineering techniques in ChatGPT to improve review comments needing further explanations. Given the recent raise of capabilities of general-purpose LLMs and their applicability to software-related tasks, we expect more and more code review automation techniques to rely on LLMs' prompt engineering rather than on fine-tuned models.

\subsubsection{Granularity} We only discuss the observed trends for a selection of the tasks, mostly those targeted by several works. For some tasks, the granularity of the targeted entities is rather homogeneous. For example, when \emph{recommending reviewers}, all 36 works take as input a code change to review, which could be a commit, a patch, or a PR. Still, the overall idea is: given a change to review, suggest the best-suited reviewers. The same observation can be made for the five works \emph{predicting pull request/code review completion time}'', and for the four \emph{visualizing code changes}. 

When looking at generative tasks, instead, differences can be observed. The most interesting ones are those related to techniques \emph{generating review comments} (11) and \emph{implementing the code change requested by a reviewer} (11). For both of them, we can see that there are three families of techniques working on (i) entire code changes, (ii) specific diff hunks, and (iii) a single method/function. Targeting these two tasks at these different granularities entails completely different levels of difficulty. Let us discuss this point for the approaches supporting the implementation of a code change required by a reviewer. Approaches working on diff elements (either an entire diff or a diff hunk) \cite{li:fse2022,huq:ist2022,lu:apsec2023,sghaier:fse2024,froemmgen:icse2024} require the ability of the approach to ``understand'' the reviewer's comment in the context of complex diff changes which could possibly span across different code elements (\eg multiple functions or even files involved). Thus, addressing these comments may be challenging, requiring modifications to several code elements.  Differently, when isolating single functions which had been commented by a reviewer in the context of a larger code change (\ie the single function may be only one of the impacted code elements) \cite{tufano:icse2021,tufano:icse2022,zhang:ase2022,lu:issre2023}, the approach has a much more limited coding context on which to operate the required code transformations. Obviously, also the applicability and potential usefulness of these techniques is different, with the former being more flexible (\eg several reviewer comments do not even concern methods, but other code elements). The recent trend is to expand as much as possible the ``contextual information'' available to these code review automation techniques. This is mostly possible thanks to the availability of large DL models able to process large inputs (such as an entire diff). This is an interesting example of how technical constraints (\eg DL models only able to process up to 512 tokes as input \cite{tufano:icse2022}) pushed researchers to artificially simplify the tackled problem (\ie only focusing on changes required to small methods \cite{tufano:icse2022}), with the most recent approaches relaxing these constraints thanks to the advances in AI.

\subsubsection{Language} Works automating code-review tasks only requiring the processing of natural language information (\ie the review comments, the pull request description) are, by definition, programming language-independent (\ie \emph{identifying/improving review comments needing further explanations}, \emph{configuring static code analysis tools}, and all those related to \emph{code review sentiment analysis} and to \emph{time management} --- see \tabref{tab:rq2}). Also, 3/4 of the works \emph{classifying the usefulness of review comments} are language independent, while one is focused on comments related to Python code. As expected, these techniques have been experimented on English artifacts only.

Among the 36 techniques \emph{recommending reviewers}, 31 are also language-independent. Indeed, many of them mostly exploit historical information and look at the source code as a bag-of-words, do not really requiring parsing or other language-specific implementations. The remaining 5 either support Java only \cite{balachandran:icse2013,sulun:icpmdase2019,ahasanuzzaman:emse2024} or a set of multiple languages \cite{rahman:icse2016,yu:ist2016}.

When looking at the remaining 78 approaches, the most targeted programming language is Java (58 works), with 35 focusing exclusively on it and 23 also supporting at least another language. For example, Li \etal \cite{li:fse2022} addressed the tasks of \emph{implementing the code change requested by a reviewer}, \emph{generating review comments}, and \emph{predicting code changes approval, merge, or need for review} by training a transformer model on code review instances related to code written in nine different languages: C, C++, C\#, Go, Java, JavaScript, PHP, Python, and Ruby. Finally, 9 of these 78 techniques are language-independent (\ie they can be applied independently from the programming language, without any adaptation) \cite{soltanifar:esem2016,li:seke2017,hirao:esecfse2019,wang:ist2021b,rahman:esem2022,islam:ist2022,chouchen:emse2024,chouchen:tosem2024,yang:emse2024}.  

Interesting is to note the complete lack of support for low-resource languages, namely programming languages for which little training material is available (\eg Julia, Lua, R). We will discuss this point further in \secref{sub:limitations}. 

%!TEX root = ../main.tex

%%%%%%%%%%%%%%%%%%%%%%%%%%%%%%%%%%%%%%%%%%%%%

\begin{table}
\centering
\caption{Evaluation of the proposed techniques: Top-3 metrics used in the evaluation; whether a qualitative inspection of the results has been performed; and whether the approaches have been deployed in industry\vspace{-0.3cm}}
\label{tab:rq3_metrics}
    {\scriptsize
   \begin{tabular}{ p{6.5cm} | p{1.4cm}| p{1.4cm} | p{1.4cm} | >{\centering\arraybackslash}p{0.9cm} | >{\centering\arraybackslash}p{0.9cm}}
    \rulec
  {\bf Task} & {\bf \#1 Metric} & {\bf \#2 Metric} & {\bf \#3 Metric} & {\bf Qualitative} & {\bf Deployed}\\  \rulec

     %ASSESSING REVIEW QUALITY ---------------------------------------------------------------------------|
      \multicolumn{5}{c}{\bf  Assessing Review Quality} \\ \rulec
     
      Assessing Review Quality through Biometrics (2) 
     & Accuracy
     & F1-score
     & Precision\&Recall
     & \xmark
     & \xmark\\
     
     \cellcolor{lightlightgrey}Classifying the Usefulness of Review Comments (4) 
     &\cellcolor{lightlightgrey} Precision\&Recall
     &\cellcolor{lightlightgrey} F1-score 
     &\cellcolor{lightlightgrey} Accuracy
     &\cellcolor{lightlightgrey} 2/4
     &\cellcolor{lightlightgrey} 2/3 \\
     
     Identifying/Improving Review Comments Needing Further Explanations(2) 
     & Accuracy
     & F1-score
     & Correct Type
     & 1/2
     & \xmark \\
     \rulec
    
    % CODE CHANGE ANALYSIS ---------------------------------------------------------------------------|
    \multicolumn{5}{c}{\bf Code Change Analysis} \\ \rulec

     \cellcolor{lightlightgrey}Decomposing Tangled Commit (3)
     &\cellcolor{lightlightgrey} Accuracy
     &\cellcolor{lightlightgrey} MAP
     &\cellcolor{lightlightgrey} MRR
     &\cellcolor{lightlightgrey} 3/3
     &\cellcolor{lightlightgrey} \raisebox{-0\height}{\strut \xmark}\\
     
     Impact Analysis for Code Review (1)
     & Accuracy
     & Recall
     & MAP 
     & 1/1 
     & \xmark\\
     
     \cellcolor{lightlightgrey}Linking Similar Contributions (2) 
     &\cellcolor{lightlightgrey} F1-score
     &\cellcolor{lightlightgrey} MRR
     &\cellcolor{lightlightgrey} Precision\&Recall
     &\cellcolor{lightlightgrey} \xmark
     &\cellcolor{lightlightgrey} \raisebox{-0\height}{\strut \xmark}\\
     
    Predicting Salient-Class (1)
     & Accuracy
     & Precision\&Recall
     & -
     & 1/1
     & \xmark\\
    \rulec

     % CODE CHANGE CLASSIFICATION ---------------------------------------------------------------------------|
    \multicolumn{5}{c}{\bf Code Change Classification} \\ \rulec
      
     \cellcolor{lightlightgrey}Identifying Impactful Code Changes (2)
     &\cellcolor{lightlightgrey} AUC
     &\cellcolor{lightlightgrey} F1-score
     &\cellcolor{lightlightgrey} Precision\&Recall
     &\cellcolor{lightlightgrey} 1/2
     &\cellcolor{lightlightgrey} 1/2\\
    
    Identifying Large-review-effort Code Changes (1) 
     & AUC
     & F1-score
     & Precision\&Recall
     & 1/1
     & 1/1\\

     \cellcolor{lightlightgrey}Identifying Quickly Reviewable Changes (1) 
     &\cellcolor{lightlightgrey} NDCG
     &\cellcolor{lightlightgrey} -
     &\cellcolor{lightlightgrey} -
     &\cellcolor{lightlightgrey} 1/1
     &\cellcolor{lightlightgrey} \xmark\\
     
     Predicting Code Changes Approval, Merge, or Need for review (11) 
     & F1-score
     & Precision\&Recall
     & AUC
     & 1/11
     & \xmark\\
    \rulec

       % CODE CHANGE QUALITY CHECK ---------------------------------------------------------------------------|
     \multicolumn{5}{c}{\bf Code Change Quality Check} \\ \rulec
     
     \cellcolor{lightlightgrey}Checking Design Patterns Consistency (1) 
     &\cellcolor{lightlightgrey} -
     &\cellcolor{lightlightgrey} - 
     &\cellcolor{lightlightgrey} -
     &\cellcolor{lightlightgrey} 1/1
     &\cellcolor{lightlightgrey} \xmark\\
     
      Generating Review Comments (11) 
     & BLEU
     & Accuracy
     & ROUGE-L 
     & 6/11
     & \xmark\\
     
     \cellcolor{lightlightgrey}Identifying Clone Refactoring Opportunities (1) 
     &\cellcolor{lightlightgrey} Accuracy
     &\cellcolor{lightlightgrey} F1-score
     &\cellcolor{lightlightgrey} Precision\&Recall
     &\cellcolor{lightlightgrey} 1/1
     &\cellcolor{lightlightgrey} \xmark\\
     
      Predicting Code Defectiveness (2)
     & F1-score %to check
     & AUC*
     & False alarms*
     & 1/2
     & 1/2\\

    \cellcolor{lightlightgrey}Predicting Problematic Code Elements (3) 
     &\cellcolor{lightlightgrey} AUC 
     &\cellcolor{lightlightgrey} F1-score
     &\cellcolor{lightlightgrey} Precision\&Recall
     &\cellcolor{lightlightgrey} 3/3
     &\cellcolor{lightlightgrey} \xmark\\
     
     Reviewing Code Formatting Violations (1)
     & F1-score
     & Precision\&Recall
     & Predicion Rate 
     & \xmark
     & \xmark\\
     
     \cellcolor{lightlightgrey}Reviewing via Static Analysis (1) 
     &\cellcolor{lightlightgrey} -
     &\cellcolor{lightlightgrey} -
     &\cellcolor{lightlightgrey} -
     &\cellcolor{lightlightgrey} 1/1
     &\cellcolor{lightlightgrey} 1/1\\
     \rulec

   % CODE REVIEW SENTIMENT ANALYIS ---------------------------------------------------------------------------|
   \multicolumn{5}{c}{\bf Code Review Sentiment Analysis} \\ \rulec
     
      Classifying the Sentiment of Review Comments (1) 
     & Accuracy
     & F1-score
     & Precision\&Recall
     & \xmark
     & \xmark\\
     
    \cellcolor{lightlightgrey}Identifying “Pushback” Feelings in Reviews (1) 
     &\cellcolor{lightlightgrey} Precision\&Recall
     &\cellcolor{lightlightgrey} -
     &\cellcolor{lightlightgrey} - 
     &\cellcolor{lightlightgrey} \xmark
     &\cellcolor{lightlightgrey} \xmark\\
     
     Identifying Toxic/Uncivil Code Review Comments (4)
     & F1-score
     & Precision\&Recall
     & Accuracy
     & \xmark
     & \xmark\\
     
     \multirow{2}{*}{\cellcolor{lightlightgrey}Rephrasing Toxic/Uncivil Comments (1)} % they also use as metric the score of models to detect the sentiment (e.g. SentiCR, SentiStrength-SE, ...)
     &\cellcolor{lightlightgrey} Incivility Deacrease
     &\cellcolor{lightlightgrey} Length Dissimilarity
     &\cellcolor{lightlightgrey} Semantic Similarity
     &\cellcolor{lightlightgrey} \multirow{2}{*}{1/1}
     &\cellcolor{lightlightgrey} \multirow{2}{*}{\xmark}\\
     \rulec
     
        % RETRIVIAL OF SIMILAR CR/CC ---------------------------------------------------------------------------|
   \multicolumn{5}{c}{\bf Retrieval of Similar CR/CC} \\ \rulec
     
      Augmenting Reviews (8)
     & Accuracy
     & Precision\&Recall % Precision 2, Recall 3
     & MRR % pari merito con BLEU
     & 4/8
     & 1/8\\
     
     \cellcolor{lightlightgrey}Mining Code Improvement Patterns (1) 
     &\cellcolor{lightlightgrey} Accuracy
     &\cellcolor{lightlightgrey} -
     &\cellcolor{lightlightgrey} -
     &\cellcolor{lightlightgrey} 1/1
     &\cellcolor{lightlightgrey} \xmark\\
     \rulec
     
     % REVISED CODE GENERATION ---------------------------------------------------------------------------|
   \multicolumn{5}{c}{\bf Revised Code Generation} \\ \rulec

     Implementing the Code Change Requested by a Reviewer (11) 
     & Accuracy
     & BLEU
     & CodeBLEU
     & 5/11
     &1/11\\
          
      \cellcolor{lightlightgrey}Predicting the Code Output of the Review Process (4) 
     &\cellcolor{lightlightgrey} Accuracy
     &\cellcolor{lightlightgrey} BLEU 
     &\cellcolor{lightlightgrey} Lev. Distance %Levenshtein Distance
     &\cellcolor{lightlightgrey} 2/4
     &\cellcolor{lightlightgrey} \xmark\\
      \rulec
     
      % TIME MANAGEMENT ---------------------------------------------------------------------------|
   \multicolumn{5}{c}{\bf Time Management} \\ \rulec
   
    Identifying Blocking Actors in Pull Requests (1)
     & MAE % Mean Absolute Error
     & MMRE
     & - 
     & 1/1
     & 1/1\\
   
     \cellcolor{lightlightgrey}Predicting Pull Request/Code Review Completion Time (5)
     &\cellcolor{lightlightgrey} MAE % Mean Absolute Error
     &\cellcolor{lightlightgrey} MRE % Mean Relative Error
     &\cellcolor{lightlightgrey} -
     &\cellcolor{lightlightgrey} 2/5
     &\cellcolor{lightlightgrey} 2/2\\
     
     Prioritizing Review Requests (3)
     & Accuracy*
     & AUC*
     & MAP*
     & 1/3
     & 2/3\\
     \rulec
     
      % OTHER ---------------------------------------------------------------------------|
   \multicolumn{5}{c}{\bf Other} \\ \rulec
   
     \cellcolor{lightlightgrey}Classifying the Goal of a Review Comment or the Type of Change Triggered by a Comment (3)
     &\cellcolor{lightlightgrey} \multirow{2}{*}{F1-score}
     &\cellcolor{lightlightgrey} \multirow{2}{*}{Precision\&Recall}
     &\cellcolor{lightlightgrey} \multirow{2}{*}{MCC} % Matthew’s Correlation Coefficient
     &\cellcolor{lightlightgrey} \multirow{2}{*}{1/3}
     &\cellcolor{lightlightgrey} \multirow{2}{*}{\xmark}\\

     Configuring Static Code Analysis Tools (1)
     & MAP % both Micro-Averaged Precision and Macro-Averaged Precision
     & MAR % Macro-Averaged Recall
     & Precision\&Recall
     & 1/1
     & \xmark\\
     
     \cellcolor{lightlightgrey}Partitioning Static Analysis Warnings (1)
     &\cellcolor{lightlightgrey} Review Effort
     &\cellcolor{lightlightgrey} -
     &\cellcolor{lightlightgrey} -
     &\cellcolor{lightlightgrey} \xmark
     &\cellcolor{lightlightgrey} \xmark\\
     
     Recommending Reviewers (36)
     & MRR % Mean Reciprocal Rank
     & Accuracy
     & Precision\&Recall
     & 4/36
     & 3/36\\
     
    \cellcolor{lightlightgrey} Visualizing Code Changes (4)
     &\cellcolor{lightlightgrey} -
     &\cellcolor{lightlightgrey} - 
     &\cellcolor{lightlightgrey} -
     &\cellcolor{lightlightgrey} 3/4 % 1 paper "We hope to empirically evaluate ... in the future
     &\cellcolor{lightlightgrey} 2/4 \\
     \rulec

    \end{tabular}
    }
\end{table}

\subsection{RQ$_3$: How are techniques for the automation of code related tasks empirically evaluated?}

\tabref{tab:rq3_metrics} shows, for each code review task $T_i$ automated in the literature: 

\begin{itemize}
\item \emph{The top-3 metrics used in the empirical evaluation of the techniques automating $T_i$}. For approaches not employing any quantitative metrics in their evaluation (\eg those \emph{visualizing code changes}) a dash is used to fill the metrics-related columns. Also, some tasks have been automated by very few techniques which, however, have been evaluated using disjointed sets of metrics. For example, the three techniques \emph{prioritizing review requests} all used different evaluation metrics \cite{chouchen:emse2024,yang:ase2024,saini:icse-seip2021}, not allowing to observe any trend. The same happens for the \emph{predicting code defectiveness} tasks.  In these cases, we just report the three metrics that are the most popular when also considering all other tasks. These cases are indicated in \tabref{tab:rq3_metrics} with a ``*'' attached to the respective metrics. Finally, we decided to group together precision and recall, since they were always used in combination in the set of inspected papers.

\item{Whether a manual qualitative inspection of the techniques' output has been performed}. A ``\xmark'' indicates that for none of the techniques automating $T_i$ a manual qualitative analysis of their output has been performed. Otherwise, a fraction explicitly shows for how many of them, out of the total, this has been done. 

\item \emph{Whether the proposed technique has been deployed in an industrial setting (``Approach Deployed'')}. This column must be read as the previous one. For example, out of the 36 techniques to \emph{recommend reviewers}, 3 have been deployed in industry.
\end{itemize}

Our goal with \tabref{tab:rq3_metrics} is to provide an overview of the evaluations performed in the literature. The interested reader can find the complete data (\eg the metrics used in each of the \selected papers) in our online appendix \cite{replication}. In the following, we are going to discuss visible trends, especially for tasks for which several automation solutions have been proposed.

We can observe a clear distinction in the metrics used for two clusters of tasks related to classification and generative problems. For the former (\eg \emph{classifying the usefulness of review comments}, \emph{identifying impactful code changes}, \emph{classifying the sentiment of review comments}, \emph{recommending reviewers}) well-known metrics such as precision, recall, F1-score, accuracy, and Area Under the ROC Curve (AUC) are mostly employed. When coming to generative tasks (\eg \emph{generating review comments}, \emph{implementing the code change requested by a reviewer}), researchers started borrowing evaluation metrics from the NLP field. For example, to assess whether a DL model is able to generate meaningful review comments, its output is compared against comments manually written by human reviewers for the same code under review, with metrics such as BLEU \cite{papineni:acl2002} (1st) and ROUGE-L \cite{lin:tsbo2004} (3rd). Both of them are basically textual-similarity metrics which only work under specific circumstances. For example, if the DL model points to the same quality issue identified by the human reviewer using, however, a completely different wording, these metrics are unable to reward the model for the meaningful output. Even more penalizing for the automated technique is the usage of accuracy (2nd), which considers a generated comment as correct only if it is identical to the human-written one. Similarly, when assessing the correctness of automatically generating code (\eg to address a review comment) researchers are using accuracy (1st), BLEU (2nd), and CodeBLEU \cite{ren:arxiv2020} (3rd). Accuracy indicates that the approach addressed the reviewer's comment exactly as done by a human developer (\ie all code tokens are identical). The CodeBLEU is  a version of the BLEU score meant to also capture AST-level similarity between two snippets of code (rather than merely textual similarity as done by BLEU). Also for this task, these evaluation metrics suffer of the same limitations discussed for the case of comment generation. Indeed, the same reviewer's comment may be successfully implemented in two different ways by the machine and by the human, with the result of low evaluation scores even in case of meaningful recommendation. We will further discuss these concerns in \secref{sub:limitations}.

Looking at \tabref{tab:rq3_metrics} it is also possible to see that in several cases researchers tried to compensate the lacks of the metrics employed to assess the effectiveness of techniques for generative problems. Indeed, 6/11 approaches \emph{generating review comments} and 5/11 of those \emph{implementing the code change requested by a reviewer} present some qualitative analysis in which, \eg researchers looked at successful and wrong recommendations with the goal of better understanding strengths and weaknesses of the proposed approaches. In general, qualitative analysis is quite popular, with 43\% of the code review automation techniques presenting some form of manual inspection. This percentage is negatively affected by the only 4/36 papers \emph{recommending reviewers} which present a qualitative analysis. 

Finally, only 18 of the \selected techniques (15\%) have been deployed in industry. For example, Froemmgen \etal \cite{froemmgen:icse2024} deployed their approach for \emph{implementing the code change requested by a reviewer} at Google. While this percentage may look low at a first sight, it is actually notable considering how recent several of the technologies behind these techniques are.

\subsection{RQ$_4$: Which techniques and datasets are publicly available?}

%!TEX root = ../main.tex

%%%%%%%%%%%%%%%%%
%%%%%%%%%%%%%%%%%
%%%%%  Table no RP  %%%%%
%%%%%%%%%%%%%%%%%
%%%%%%%%%%%%%%%%%

\begin{table}
	\centering
    \caption{Works on code review automation not providing a replication package or having it not accessible as of Jan 2025\vspace{-0.3cm}}
    \label{tab:replicationNo}
    {\scriptsize
   \begin{tabular}{p{7.0cm} | p{2.8cm} | c | c } 
    \rulec
    {\bf Task} & {\bf Reference} & {\bf Provided} & {\bf Accessible}\\
    \rulec

%%%%%

Assessing Review Quality through Biometrics &\cellcolor{lightlightgrey}Hijazi \etal \cite{hijazi:issre2021} &\cellcolor{lightlightgrey}\xmark & \cellcolor{lightlightgrey}- \\[0.05cm]\hline

%%%%%

\multirow{5}{*}{Augmenting Reviews} &Guo \etal \cite{guo:saner2019} &\xmark &- \\[0.05cm]

&\cellcolor{lightlightgrey}Guo \etal \cite{guo2020review} &\cellcolor{lightlightgrey}\xmark &\cellcolor{lightlightgrey}- \\[0.05cm]

&Gupta \etal \cite{gupta:sigkdd2018} &\xmark & - \\[0.05cm]

&\cellcolor{lightlightgrey}Rahman \etal \cite{rahman:esem2022} &\cellcolor{lightlightgrey}\xmark &\cellcolor{lightlightgrey}- \\[0.05cm]\hline

%%%%%

Checking Design Patterns Consistency &He \etal \cite{he:icssrc2013} &\xmark &- \\[0.05cm]\hline

%%%%%

\multirow{2}{*}{Classifying the Usefulness of Review Comments} &\cellcolor{lightlightgrey}Pangsakulyanont \etal \cite{pangsakulyanont:iwesep2014} &\cellcolor{lightlightgrey}\xmark & \cellcolor{lightlightgrey}- \\

&Rahman \etal \cite{rahman:msr2017} &\cmark &\xmark \\[0.05cm]\hline

%%%%%

\multirow{3}{*}{Decomposing Tangled Commit} &\cellcolor{lightlightgrey}Barnett \etal \cite{barnett:icse2015} &\cellcolor{lightlightgrey}\xmark & \cellcolor{lightlightgrey}- \\

&Tao \etal \cite{tao:msr2015} &\xmark &- \\[0.05cm]

&\cellcolor{lightlightgrey}Wang \etal \cite{wang:ase2019} &\cellcolor{lightlightgrey}\xmark &\cellcolor{lightlightgrey}- \\[0.05cm]\hline

%%%%%

\multirow{2}{*}{Generating Review Comments} & Nashaat \etal \cite{nashaat:tse2024} &\xmark & - \\[0.05cm]

&\cellcolor{lightlightgrey}Vijayvergiya \etal \cite{vijayvergiya:aiware2024} &\cellcolor{lightlightgrey}\xmark &\cellcolor{lightlightgrey}- \\[0.05cm]\hline

%%%%%

Identifying Blocking Actors in Pull Requests &Shan \etal \cite{shan:esecfse2022} &\xmark &- \\[0.05cm]\hline

%%%%%

Identifying Clone Refactoring Opportunities &\cellcolor{lightlightgrey}Chen \etal \cite{chen:compsac2017} &\cellcolor{lightlightgrey}\cmark & \cellcolor{lightlightgrey}\xmark \\[0.05cm]\hline

%%%%%

Identifying Impactful Code Changes &Wen \etal \cite{wen:icsme2018} &\xmark & - \\[0.05cm]\hline

%%%%%

Identifying Quickly Reviewable Changes &\cellcolor{lightlightgrey}Zhao \etal \cite{zhao:emse2019} &\cellcolor{lightlightgrey}\xmark &\cellcolor{lightlightgrey}- \\[0.05cm]\hline

%%%%%

Identifying ``Pushback'' Feelings in Reviews &Egelman \etal \cite{egelman:icse2020} &\xmark &- \\[0.05cm]\hline

%%%%%

Identifying/Improving Review Comments Needing Further Explanations &\cellcolor{lightlightgrey}Rahman \etal \cite{rahman:esem2022} &\cellcolor{lightlightgrey}\xmark &\cellcolor{lightlightgrey}- \\[0.05cm]\hline

%%%%%

\multirow{2}{*}{Implementing the Code Change Requested by a Reviewer} &Froemmgen \etal \cite{froemmgen:icse2024} &\xmark &- \\[0.05cm]

&\cellcolor{lightlightgrey}Nashaat \etal \cite{nashaat:tse2024} &\cellcolor{lightlightgrey}\xmark &\cellcolor{lightlightgrey}- \\[0.05cm]\hline

%%%%%

Linking Similar Contributions &Ayinala \etal \cite{ayinala:compsac2020} &\cmark &\xmark \\[0.05cm]\hline

%%%%%

Mining Code Improvement Patterns &\cellcolor{lightlightgrey}Ueda \etal \cite{ueda:iwsc2019} &\cellcolor{lightlightgrey}\xmark &\cellcolor{lightlightgrey}- \\[0.05cm]\hline

%%%%%

Partitioning Static Analysis Warnings &Tukaram \etal \cite{tukaram:scam2013} &\xmark & - \\[0.05cm]\hline

%%%%%

\multirow{3}{*}{Predicting Code Changes Approval, Merge, or Need for review} &\cellcolor{lightlightgrey}Fan \etal \cite{fan:emse2018} &\cellcolor{lightlightgrey}\xmark & \cellcolor{lightlightgrey}- \\

&Nashaat \etal \cite{nashaat:tse2024} &\xmark &- \\[0.05cm]

&\cellcolor{lightlightgrey}Shi \etal \cite{shi:2019} &\cellcolor{lightlightgrey}\xmark &\cellcolor{lightlightgrey}- \\[0.05cm]

\hline

%%%%%

\multirow{2}{*}{Predicting Code Defectiveness} &Sharma \etal \cite{sharma:spe2019} &\xmark &- \\[0.05cm]

&\cellcolor{lightlightgrey}Soltanifar \etal \cite{soltanifar:esem2016} &\cellcolor{lightlightgrey}\xmark & \cellcolor{lightlightgrey}- \\

\hline

%%%%%

\multirow{3}{*}{Predicting Pull Requests/Code Review Completion Time} &Chen \etal \cite{chen:esecfse2022} &\xmark & - \\[0.05cm]

&\cellcolor{lightlightgrey}Maddila \etal \cite{maddila:esefse2019} &\cellcolor{lightlightgrey}\xmark & \cellcolor{lightlightgrey}- \\

&Shan \etal \cite{shan:esecfse2022} &\xmark & - \\[0.05cm]
\hline

%%%%%

Predicting Salient-Class &\cellcolor{lightlightgrey}Huang \etal \cite{huang:tse2020} &\cellcolor{lightlightgrey}\xmark &\cellcolor{lightlightgrey}- \\[0.05cm]\hline

%%%%%

Prioritizing Review Requests &Saini \etal \cite{saini:icse-seip2021} &\xmark & - \\[0.05cm]\hline

%%%%%

\multirow{23}{*}{Recommending Reviewers} &\cellcolor{lightlightgrey}Al \etal \cite{al:icpmda2020} &\cellcolor{lightlightgrey}\xmark &\cellcolor{lightlightgrey}- \\[0.05cm]

&Aryendu \etal \cite{aryendu:ase2023} &\xmark &- \\[0.05cm]

&\cellcolor{lightlightgrey}Asthana \etal \cite{asthana:esecfse2019} &\cellcolor{lightlightgrey}\xmark &\cellcolor{lightlightgrey}- \\[0.05cm]

&Balachandran \etal \cite{balachandran:icse2013} &\xmark & - \\

&\cellcolor{lightlightgrey}Chouchen \etal \cite{chouchen:asc2021} &\cellcolor{lightlightgrey}\xmark &\cellcolor{lightlightgrey}- \\[0.05cm]

&Jiang \etal \cite{jiang:jcst2015} &\xmark & - \\[0.05cm]

&\cellcolor{lightlightgrey}Jiang \etal \cite{jiang:jss2019} &\cellcolor{lightlightgrey}\xmark &\cellcolor{lightlightgrey}- \\[0.05cm]

&Jiang \etal \cite{jiang:ist2017} &\xmark &- \\[0.05cm]

&\cellcolor{lightlightgrey}Kong \etal \cite{kong:saner2022} &\cellcolor{lightlightgrey}\xmark & \cellcolor{lightlightgrey}- \\[0.05cm]

&Liao \etal \cite{liao:globecom2019} &\xmark &- \\[0.05cm]

&\cellcolor{lightlightgrey}Ouni \etal \cite{ouni:icsme2016} &\cellcolor{lightlightgrey}\xmark &\cellcolor{lightlightgrey}- \\[0.05cm]

&Pandya \etal \cite{pandya:esecfse2022} &\cmark &\xmark \\[0.05cm]

&\cellcolor{lightlightgrey}Rahman \etal \cite{rahman:icse2016} &\cellcolor{lightlightgrey}\cmark &\cellcolor{lightlightgrey}\xmark \\[0.05cm]

&Rebai \etal \cite{rebai:ase2020} &\xmark &- \\[0.05cm]

&\cellcolor{lightlightgrey}Rong \etal \cite{rong:icse2022} &\cellcolor{lightlightgrey}\xmark & \cellcolor{lightlightgrey}- \\[0.05cm]

&Strand \etal \cite{strand:icse-sep2020} &\xmark &- \\[0.05cm]

&\cellcolor{lightlightgrey}Xia \etal \cite{xia:icsme2015} &\cellcolor{lightlightgrey}\xmark &\cellcolor{lightlightgrey}- \\[0.05cm]

&Xia \etal \cite{xia:sm2017} &\xmark &- \\[0.05cm]

&\cellcolor{lightlightgrey}Ye \etal \cite{ye:ieee2019} &\cellcolor{lightlightgrey}\xmark & \cellcolor{lightlightgrey}- \\[0.05cm]

&Ying \etal \cite{ying:csi2016} &\xmark &- \\[0.05cm]

&\cellcolor{lightlightgrey}Yu \etal \cite{yu:ist2016} &\cellcolor{lightlightgrey}\xmark &\cellcolor{lightlightgrey}- \\[0.05cm]

&Zanjan \etal \cite{zanjan:tse2016} &\cmark &\xmark \\[0.05cm]

&\cellcolor{lightlightgrey}Zhang \etal \cite{zhang:icse2023} &\cellcolor{lightlightgrey}\xmark &\cellcolor{lightlightgrey}- \\[0.05cm]\hline

%%%%%

Reviewing via Static Analysis &Balachandran \etal \cite{balachandran:icse2013} &\xmark & - \\[0.05cm]\hline

%%%%%

Visualizing Code Changes &\cellcolor{lightlightgrey}Menarini \etal \cite{menarini:ase2017} &\cellcolor{lightlightgrey}\xmark &\cellcolor{lightlightgrey}- \\[0.05cm]

    \rulec
    \end{tabular}
    }
\end{table}

%!TEX root = ../main.tex

%%%%%%%%%%%%%%%%%
%%%%%%%%%%%%%%%%%
%%%%%  Table RP  %%%%%
%%%%%%%%%%%%%%%%%
%%%%%%%%%%%%%%%%%

\begin{table}
	\centering
    \caption{Works on code review automation providing a (still accessible at Jan 2025) replication package \vspace{-0.4cm}}
    \label{tab:replication}
    {\scriptsize
   \begin{tabular}{p{4.7cm} | p{2.5cm} | l | c | c} 
    \rulec
    {\bf Task} & {\bf Reference} & {\bf Link} & {\bf C} & {\bf D}\\
    \rulec

\multirow{3}{*}{Augmenting Reviews} & Hirao \etal \cite{hirao:esecfse2019} & \url{https://github.com/software-rebels/ReviewLinkageGraph} & \cmark & \cmark \\[0.05cm]

& \cellcolor{lightlightgrey}Kartal \etal \cite{kartal:infsof2024} & \cellcolor{lightlightgrey}\url{https://github.com/ykartal/Github-SourceCode-Review} & \cellcolor{lightlightgrey}\cmark & \cellcolor{lightlightgrey}\cmark \\[0.05cm]

& Shuvo \etal \cite{shuvo:icsme2023} & \url{https://drive.google.com/file/d/15kq7LqvfY-oP1M1UDdK_lLmUfq71daVR/view} & \xmark & \cmark \\[0.05cm]
\hline

%%%%%

Classifying the Goal of a Review Comment & \cellcolor{lightlightgrey}Fregnan \etal \cite{fregnan:emse2022} & \cellcolor{lightlightgrey}\url{https://zenodo.org/records/5592254} & \cellcolor{lightlightgrey}\cmark & \cellcolor{lightlightgrey}\cmark \\[0.05cm]

or the Type of Change Triggered by a Comment & Li \etal \cite{li:seke2017} & \url{https://sites.google.com/view/core2019/} &\xmark &\cmark \\[0.05cm]

& \cellcolor{lightlightgrey}Turzo \etal \cite{Turzo:esem2023} & \cellcolor{lightlightgrey}\url{https://github.com/WSU-SEAL/CR-classification-ESEM23} & \cellcolor{lightlightgrey}\cmark & \cellcolor{lightlightgrey}\cmark \\[0.05cm] \hline

%%%%%

\multirow{2}{*}{Classifying the Usefulness of Review Comments} & Hasan \etal \cite{hasan:emse2021} &\url{https://github.com/WSU-SEAL/CRA-usefulness-model} &\cmark &\cmark \\[0.05cm]

& \cellcolor{lightlightgrey}Yang \etal \cite{yang:fse2023} &\cellcolor{lightlightgrey}\url{https://zenodo.org/records/8297481} &\cellcolor{lightlightgrey}\cmark &\cellcolor{lightlightgrey}\cmark \\[0.05cm]

\hline

%%%%%

\multirow{10}{*}{Generating Review Comments} &Hong \etal \cite{hong:esecfse2022} &\url{https://github.com/awsm-research/CommentFinder} &\cmark &\cmark \\[0.05cm]

&\cellcolor{lightlightgrey}Li \etal \cite{li:fse2022} &\cellcolor{lightlightgrey}\url{https://github.com/microsoft/CodeBERT/tree/master/CodeReviewer} &\cellcolor{lightlightgrey}\cmark &\cellcolor{lightlightgrey}\cmark \\[0.05cm]

&Li \etal \cite{li.l:esecfse2022} &\url{https://gitlab.com/ai-for-se-public-data/auger-fse-2022} &\cmark &\cmark \\[0.05cm]

&\cellcolor{lightlightgrey}Lin \etal \cite{lin:msr2024} &\cellcolor{lightlightgrey}\url{https://zenodo.org/records/10572047} &\cellcolor{lightlightgrey}\cmark &\cellcolor{lightlightgrey}\cmark \\[0.05cm]

&Lu \etal \cite{lu:issre2023} &\url{https://zenodo.org/records/7991113} &\cmark &\cmark \\[0.05cm]

&\cellcolor{lightlightgrey}Lu \etal \cite{lu:ase2024} &\cellcolor{lightlightgrey}\url{https://zenodo.org/records/10964945} &\cellcolor{lightlightgrey}\cmark &\cellcolor{lightlightgrey}\cmark \\[0.05cm]

&Sghaier \etal \cite{sghaier:fse2024} &\url{https://zenodo.org/records/10676741} &\cmark &\cmark \\[0.05cm]

&\cellcolor{lightlightgrey}Tufano \etal \cite{tufano:icse2022} &\cellcolor{lightlightgrey}\url{https://github.com/RosaliaTufano/code_review_automation} &\cellcolor{lightlightgrey}\cmark &\cellcolor{lightlightgrey}\cmark \\

&Yu \etal \cite{yu:tosem2024} &\url{https://github.com/aiopsplus/Carllm} &\xmark &\cmark \\[0.05cm]\hline

%%%%%

\multirow{5}{*}{Identifying Toxic/Uncivil Code Review Comments} & \cellcolor{lightlightgrey}Ferreira \etal \cite{ferreira:jss2024} &\cellcolor{lightlightgrey}\url{https://doi.org/10.6084/m9.figshare.24603237} &\cellcolor{lightlightgrey}\cmark &\cellcolor{lightlightgrey}\cmark \\[0.05cm]

&Rahman \etal \cite{rahman:fse2024} &\url{https://github.com/Oyakiolo052/ATUC_Artifacts} &\cmark &\cmark \\[0.05cm]

&\cellcolor{lightlightgrey}Sarker \etal \cite{sarker:tosem2023} &\cellcolor{lightlightgrey}\url{https://github.com/WSU-SEAL/ToxiCR} &\cellcolor{lightlightgrey}\cmark &\cellcolor{lightlightgrey}\cmark \\[0.05cm]

&Sarker \etal \cite{sarker:esem2023} &\url{https://github.com/WSU-SEAL/ToxiSpanSE} &\cmark &\cmark \\[0.05cm]\hline

%%%%%

Identifying/Improve Review Comments Needing Further Explanations & \cellcolor{lightlightgrey}Widyasari \etal \cite{widyasari:tosem2024} & \cellcolor{lightlightgrey}\url{https://figshare.com/s/135201b8f87ab705448b} &\cellcolor{lightlightgrey}\cmark &\cellcolor{lightlightgrey}\cmark \\[0.05cm]\hline

%%%%%

Impact Analysis for Code Review & Hong \etal \cite{hong:ist2024} & \url{https://figshare.com/s/135201b8f87ab705448b} &\cmark &\cmark \\[0.05cm]\hline

%%%%%

 & \cellcolor{lightlightgrey}Huq \etal \cite{huq:ist2022} &\cellcolor{lightlightgrey}\url{https://github.com/Review4Repair/Review4Repair} &\cellcolor{lightlightgrey}\cmark &\cellcolor{lightlightgrey}\cmark \\[0.05cm]

&Li \etal \cite{li:fse2022} &\url{https://github.com/microsoft/CodeBERT/tree/master/CodeReviewer} &\cmark &\cmark \\[0.05cm]

&\cellcolor{lightlightgrey}Lu \etal \cite{lu:apsec2023} &\cellcolor{lightlightgrey}\url{https://github.com/moonmengmeng/EnRefiner} &\cellcolor{lightlightgrey}\cmark &\cellcolor{lightlightgrey}\cmark \\[0.05cm]

Implementing the Code Change Requested by &Lu \etal \cite{lu:issre2023} &\url{https://zenodo.org/records/7991113} &\cmark &\cmark \\[0.05cm]

a Reviewer&\cellcolor{lightlightgrey}Pornprasit \etal \cite{pornprasit:ist2024} &\cellcolor{lightlightgrey}\url{https://github.com/awsm-research/LLM-for-code-review-automatiton} &\cellcolor{lightlightgrey}\cmark &\cellcolor{lightlightgrey}\cmark \\[0.05cm]

&Sghaier \etal \cite{sghaier:fse2024} &\url{https://zenodo.org/records/10676741} &\cmark &\cmark \\[0.05cm]

&\cellcolor{lightlightgrey}Tufano \etal \cite{tufano:icse2022} &\cellcolor{lightlightgrey}\url{https://github.com/RosaliaTufano/code_review_automation} &\cellcolor{lightlightgrey}\cmark &\cellcolor{lightlightgrey}\cmark \\[0.05cm]

&Zhang \etal \cite{zhang:ase2022} &\url{https://github.com/EngineeringSoftware/CoditT5} &\cmark &\cmark \\[0.05cm]\hline

%%%%%

& \cellcolor{lightlightgrey}Chouchen \etal \cite{chouchen:emse2024} &\cellcolor{lightlightgrey}\url{https://github.com/stilab-ets/CostAwareCR} &\cellcolor{lightlightgrey}\cmark &\cellcolor{lightlightgrey}\cmark \\[0.05cm]

&Chouchen \etal \cite{chouchen:tosem2024} & \url{https://github.com/stilab-ets/multicr} &\cmark &\cmark \\[0.05cm]

&\cellcolor{lightlightgrey}Islam \etal \cite{islam:ist2022} &\cellcolor{lightlightgrey}\url{https://github.com/khairulislam/Predict-Code-Changes} &\cellcolor{lightlightgrey}\cmark &\cellcolor{lightlightgrey}\cmark \\[0.05cm]

Predicting Code Changes Approval, Merge, &Li \etal \cite{li:fse2022} &\url{https://github.com/microsoft/CodeBERT/tree/master/CodeReviewer} &\cmark &\cmark \\[0.05cm]

or Need for review&\cellcolor{lightlightgrey}Lu \etal \cite{lu:issre2023} & \cellcolor{lightlightgrey}\url{https://zenodo.org/records/7991113} &\cellcolor{lightlightgrey}\cmark &\cellcolor{lightlightgrey}\cmark \\[0.05cm]

&Wu and Zhang \cite{wu2022contrastive} &\url{https://github.com/SimAST-GCN/CLMN} &\cmark &\cmark \\[0.05cm]

&\cellcolor{lightlightgrey}Wu \etal \cite{wu:kbs2022} &\cellcolor{lightlightgrey}\url{https://github.com/SimAST-GCN/SimAST-GCN} &\cellcolor{lightlightgrey}\cmark &\cellcolor{lightlightgrey}\cmark \\[0.05cm]

&Yang \etal \cite{yang:emse2024} &\url{https://figshare.com/s/7930029ea5ec5af2845d} &\cmark &\cmark \\[0.05cm]

\hline

%%%%%

\multirow{3}{*}{Predicting Problematic Code Elements} & \cellcolor{lightlightgrey}Hong \etal \cite{hong:saner2022} &\cellcolor{lightlightgrey}\url{https://github.com/awsm-research/RevSpot-replication-package} &\cellcolor{lightlightgrey}\cmark &\cellcolor{lightlightgrey}\cmark \\[0.05cm]

&Olewicki \etal \cite{olewicki:fse2024} & \url{https://zenodo.org/records/10783562} &\cmark &\cmark \\[0.05cm]

&\cellcolor{lightlightgrey}Sghaier \etal \cite{sghaier:saner2023} & \cellcolor{lightlightgrey}\url{https://zenodo.org/records/7533156} &\cellcolor{lightlightgrey}\cmark &\cellcolor{lightlightgrey}\cmark \\[0.05cm]

\hline

%%%%%

\multirow{2}{*}{Predicting Pull Request/Code Review Completion Time} &Chouchen \etal \cite{chouchen:emse2023} & \url{https://github.com/stilab-ets/MCRDuration} &\cmark &\cmark \\[0.05cm]

&\cellcolor{lightlightgrey}Yang \etal \cite{yang:emse2024} & \cellcolor{lightlightgrey}\url{https://zenodo.org/records/7533156} &\cellcolor{lightlightgrey}\cmark &\cellcolor{lightlightgrey}\cmark \\[0.05cm]

\hline

%%%%%

Predicting the Code Output of the Review Process & Pornprasit \etal \cite{pornprasit:saner2023} & \url{https://github.com/awsm-research/D-ACT-Replication-Package} &\cmark &\cmark \\[0.05cm]\hline

%%%%%

\multirow{2}{*}{Prioritizing Review Requests} & \cellcolor{lightlightgrey}Chouchen \etal \cite{chouchen:emse2024} & \cellcolor{lightlightgrey}\url{https://github.com/stilab-ets/CostAwareCR} &\cellcolor{lightlightgrey}\cmark &\cellcolor{lightlightgrey}\cmark \\[0.05cm]

& Yang \etal \cite{yang:ase2024} & \url{https://figshare.com/s/133f23da558b7b254041?file=46923235} &\cmark &\cmark \\[0.05cm] 

\hline

%%%%%

\multirow{15}{*}{Recommending Reviewers} &\cellcolor{lightlightgrey}Ahasanuzzaman \etal \cite{ahasanuzzaman:emse2024} &\cellcolor{lightlightgrey}\url{https://drive.google.com/drive/folders/1bSC9iRtjKjMTRa9hiyECijgABKGfpyT4} &\cellcolor{lightlightgrey}\cmark &\cellcolor{lightlightgrey}\cmark \\[0.05cm]

&Chueshev \etal \cite{chueshev:icsme2020} &\url{https://github.com/alexchueshev/icsme2020} &\cmark &\cmark \\[0.05cm]

&\cellcolor{lightlightgrey}Fejzer \etal \cite{fejzer:jiis2018} &\cellcolor{lightlightgrey}\url{https://github.com/mfejzer/reviewers_recommendation} &\cellcolor{lightlightgrey}\cmark &\cellcolor{lightlightgrey}\cmark \\[0.05cm]

&Hajari \etal \cite{hajari:tse2024} &\url{https://github.com/rigbypc/SofiaWL/tree/master/ReplicationPackage} &\cmark &\cmark \\[0.05cm]

&\cellcolor{lightlightgrey}Li \etal \cite{li:ease2023} &\cellcolor{lightlightgrey}\url{https://zenodo.org/record/7292881} &\cellcolor{lightlightgrey}\cmark &\cellcolor{lightlightgrey}\cmark \\[0.05cm]

&Mirsaeedi \etal \cite{mirsaeedi:icse2020} &\url{https://zenodo.org/record/3678551#.ZFS5EC8RpBw} &\xmark &\cmark \\[0.05cm]

&\cellcolor{lightlightgrey}Qiao \etal \cite{qiao:saner2024} &\cellcolor{lightlightgrey}\url{https://github.com/cufeinfor/MIRRec} &\cellcolor{lightlightgrey}\cmark &\cellcolor{lightlightgrey}\cmark \\[0.05cm]

&Rahman \etal \cite{rahman:icsme2023} &\url{https://zenodo.org/records/8190493} &\cmark &\cmark \\[0.05cm]

&\cellcolor{lightlightgrey}Sulun \etal \cite{sulun:icpmdase2019} &\cellcolor{lightlightgrey}\url{https://figshare.com/s/27a35b4ae70269481a2c} &\cellcolor{lightlightgrey}\cmark &\cellcolor{lightlightgrey}\cmark \\[0.05cm]

&Sulun \etal \cite{sulun:ist2021} &\url{https://github.com/sulunemre/rstrace-replication} &\cmark &\cmark \\[0.05cm]

&\cellcolor{lightlightgrey}Tecimer \etal \cite{tecimer:ease2021} &\cellcolor{lightlightgrey}\url{https://figshare.com/s/1b9ea55377d9f2c31a7a} &\cellcolor{lightlightgrey}\cmark &\cellcolor{lightlightgrey}\cmark \\[0.05cm]

&Thongtanunam \etal \cite{thongtanunam:saner2015} &\url{https://github.com/patanamon/revfinder} &\xmark &\cmark \\[0.05cm]

&\cellcolor{lightlightgrey}Zhao \etal \cite{zhao:cascon:2022} &\cellcolor{lightlightgrey}\url{https://github.com/liuj888/ReviewerRecommendationLtR} &\cellcolor{lightlightgrey}\xmark &\cellcolor{lightlightgrey}\xmark \\[0.05cm]

\rulec
    \end{tabular}
    }
\end{table}

\begin{table}[h!]
	\centering
    \caption*{Table 9 (continue): Works on code review automation providing a (still accessible at Jan 2025) replication package \vspace{-0.4cm}}
    {\scriptsize
   \begin{tabular}{p{4.7cm} | p{2.5cm} | l | c | c} 
    \rulec
    {\bf Task} & {\bf Reference} & {\bf Link} & {\bf C} & {\bf D}\\
    \rulec

%%%%%

Rephrasing Toxic/Uncivil Comments & Rahman \etal \cite{rahman:fse2024} & \url{https://github.com/Oyakiolo052/ATUC_Artifacts} &\cmark &\cmark \\[0.05cm]\hline

%%%%%

Linking Similar Contributions & \cellcolor{lightlightgrey}Wang \etal \cite{wang:ist2021a} & \cellcolor{lightlightgrey}\url{https://github.com/dong-w/Replication-Patch-Linkage} & \cellcolor{lightlightgrey}\cmark & \cellcolor{lightlightgrey}\cmark \\[0.05cm]\hline

%%%%%

Identifying Impactful Code Changes & Uch\^oa \etal \cite{uchoa:msr2021} &\url{https://zenodo.org/record/4563214#.Y0kjiexBwQg} &\xmark &\cmark \\[0.05cm]\hline

%%%%%

Identifying Large-review-effort Code Changes & \cellcolor{lightlightgrey}Wang \etal \cite{wang:ist2021b} & \cellcolor{lightlightgrey}\url{https://bitbucket.org/wangsonging/ist2020_repo/src/master/} &\cellcolor{lightlightgrey}\cmark &\cellcolor{lightlightgrey}\cmark \\[0.05cm]\hline 

%%%%%

Reviewing Code Formatting Violations & Markovtsev \etal \cite{markovtsev:msr2019} & \url{https://github.com/src-d/style-analyzer} &\cmark &\cmark \\[0.05cm]\hline 

%%%%%

Assessing Review Quality through Biometrics & \cellcolor{lightlightgrey}Hijazi \etal \cite{hijazi:tse2022} & \cellcolor{lightlightgrey}\url{https://github.com/HaythamHijazi/Supplement} &\cellcolor{lightlightgrey}\cmark &\cellcolor{lightlightgrey}\cmark \\[0.05cm]\hline 

%%%%%

Classifying the Sentiment of Review Comments & Ahmed \etal \cite{ahmed:ase2017} &\url{https://github.com/senticr/SentiCR/} &\cmark &\cmark \\[0.05cm]\hline

%%%%%

Retrieving Similar Reviews & \cellcolor{lightlightgrey}Siow \etal \cite{siow:saner2020} & \cellcolor{lightlightgrey}\url{https://sites.google.com/view/core2019/} &\cellcolor{lightlightgrey}\xmark &\cellcolor{lightlightgrey}\cmark \\[0.05cm]\hline 

%%%%%

\multirow{5}{*}{Predicting the Code Output of the Review Process}  &Huq \etal \cite{huq:ist2022} &\url{https://github.com/Review4Repair/Review4Repair} &\cmark &\cmark \\[0.05cm]

&\cellcolor{lightlightgrey}Patanamon \etal \cite{patanamon:icse2022} &\cellcolor{lightlightgrey}\url{https://github.com/awsm-research/AutoTransform-Replication} &\cellcolor{lightlightgrey}\cmark &\cellcolor{lightlightgrey}\cmark \\[0.05cm]

&Tufano \etal \cite{tufano:icse2021} &\url{https://github.com/RosaliaTufano/code_review} &\cmark &\cmark \\[0.05cm]

&\cellcolor{lightlightgrey}Tufano \etal \cite{tufano:icse2022} &\cellcolor{lightlightgrey}\url{https://github.com/RosaliaTufano/code_review_automation} &\cellcolor{lightlightgrey}\cmark &\cellcolor{lightlightgrey}\cmark \\[0.05cm]

\hline

%%%%%

\multirow{3}{*}{Visualizing Code Changes} &Brito and Valente \cite{brito:icpc2021} &\url{https://github.com/rodrigo-brito/refactoring-aware-diff} &\cmark &\xmark \\[0.05cm]

&\cellcolor{lightlightgrey}Fadhel \etal \cite{fadhel:iccq2021} &\cellcolor{lightlightgrey}\url{https://github.com/hadii-tech/striff-lib} &\cellcolor{lightlightgrey}\cmark &\cellcolor{lightlightgrey}\xmark \\

&Fregnan \etal \cite{fregnan:jss2023} &\url{https://zenodo.org/record/7047993#.Y2JqNS-B2Uo} &\cmark &\cmark \\[0.05cm]\hline

%%%%%

Configuring Static Code Analysis Tools & \cellcolor{lightlightgrey}Zampetti \etal \cite{zampetti:emse2022} &\cellcolor{lightlightgrey}\url{https://github.com/senticr/SentiCR/} &\cellcolor{lightlightgrey}\xmark &\cellcolor{lightlightgrey}\cmark \\[0.05cm]

    \rulec
    \end{tabular}
    }
\end{table}

\tabref{tab:replicationNo} reports the list of works from our SLR which either do not provide a link to a replication package (\xmark in column ``Provided'') or, while having such a link (\cmark), it is not accessible (\xmark~in column ``Accessible'') at the date of writing (January 2025). Some references are present multiple times since the proposed approach supports several tasks. 

Overall, 49 of the \selected papers (41\%) part of our SLR do not provide a replication package, and 6 more (5\%) provide a link which is not accessible anymore. Considering that all surveyed papers present techniques for automating code review tasks, this implies substantial challenges for researchers interested in replicating these approaches, for example to use them as baselines for the proposal of a novel solution. The remaining 64 papers (54\%) provide instead a working replication package, as documented in \tabref{tab:replication}. Besides reporting the link to the working replication package, \tabref{tab:replication} also indicates what the authors provide in it in terms of code/tool implementing the proposed approach (column ``C'') and data used in the paper (column ``D''). Note that one of the works \cite{zhao:cascon:2022} does not provide both code and data, with the linked artifact mostly presenting additional tables. The most popular platform used for sharing the replication packages is by far GitHub (61\%), followed by other solutions with a similar usage share (\ie Zenodo, Figshare, bitbucket, personal website).

\begin{figure*}[h!]
	\centering
	\includegraphics[width=0.8\linewidth]{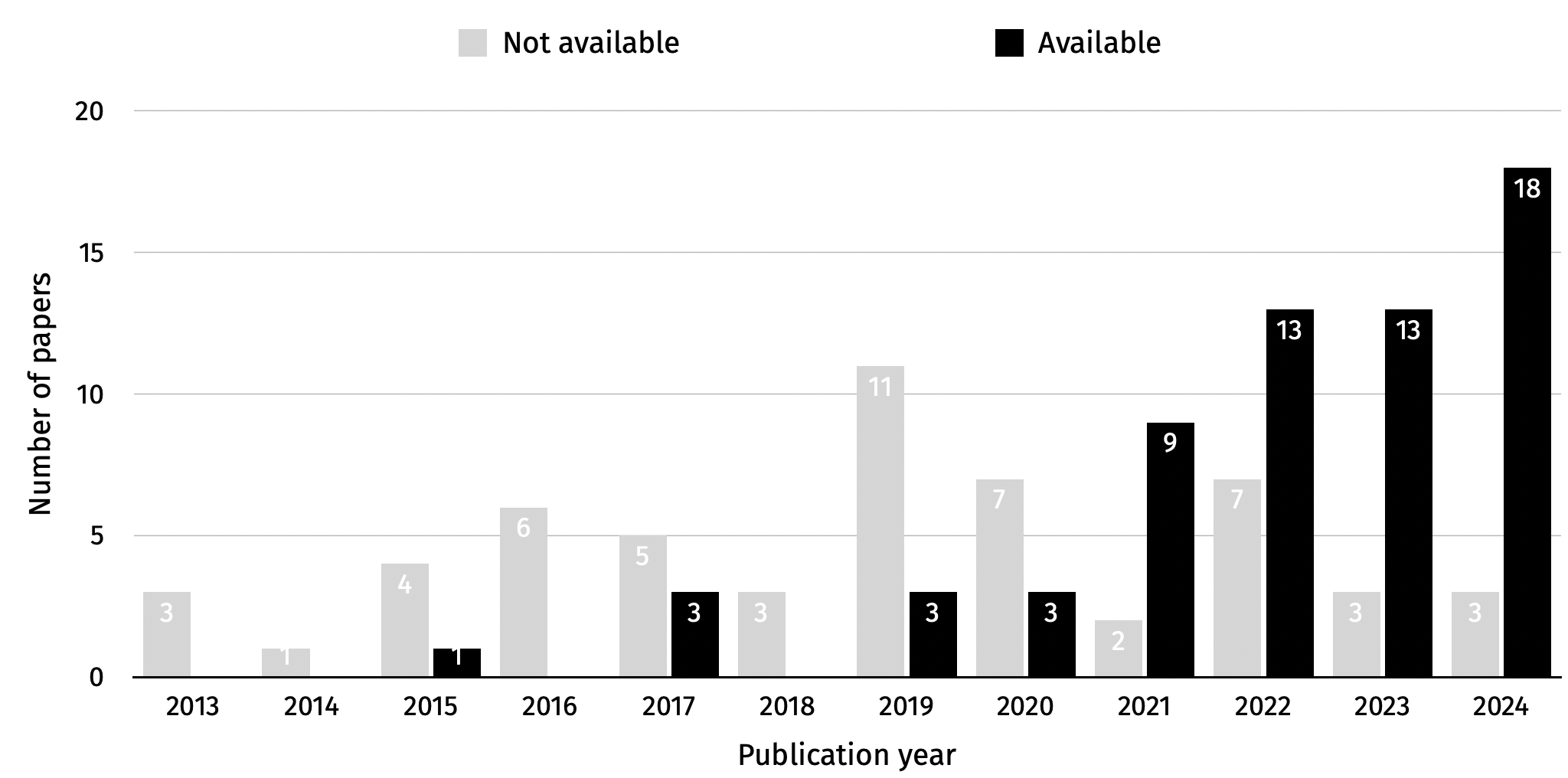}
	\caption{Availability of a working replication package by publication year}
	\label{fig:replicability}
\end{figure*}

While the percentage of papers providing a working replication package (54\%) seems to suggest major issues in the replicability of techniques for code review automation, it is important to look at how such a trend is evolving over time. \figref{fig:replicability} shows that the efforts put in place by the software engineering research community for promoting open science (\eg by default all papers submitted at the International Conference on Software Engineering must disclose data/artifacts) are improving code/data availability. 

Indeed, while up to 2020 the majority of the published papers did not provide a working replication package, such a trend changed in 2021 (81\% provide a replication package) and was confirmed in all subsequent years, with 86\% of published works providing a replication package in 2024. 

Encouraging signals also come from the information reported in \tabref{tab:replication}: Indeed, of the 64 papers providing a replication package, 54 (84\%) disclose both the source code of the proposed approach and/or a tool implementing the approach and data used in the work.

\subsection{RQ$_5$: What are the concerns raised or the limitations observed by researchers when experimenting the automated solutions?}
\label{sub:limitations}

\begin{table}
	\centering
    \caption{Concerns and limitations discussed by researchers\vspace{-0.3cm}}
    \label{tab:concerns}
    {\footnotesize
   \begin{tabular}{l | l | p{5cm}} 
    \rulec
    {\bf Parent Category} & {\bf Child Category} & {\bf References}\\
    \rulec
	\multirow{6}{*}{{\bf Performance}} & Limited support in specific scenarios & \cite{barnett:icse2015,tao:msr2015,huang:tse2020,hong:esecfse2022,tufano:icse2021,hong:ist2024,yang:fse2023,sghaier:fse2024,sulun:ist2021,nashaat:tse2024}\\
	& \cellcolor{lightlightgrey}Lack of generalizability across different datasets & \cellcolor{lightlightgrey} \cite{ferreira:jss2024,nashaat:tse2024,pangsakulyanont:iwesep2014}\\
	& Unsatisfactory Recommendations &  \cite{patanamon:icse2022,egelman:icse2020,balachandran:icse2013,huq:ist2022,wang:ist2021a,tufano:icse2021,zampetti:emse2022,tufano:icse2022,li:seke2017,petersson:issre2001,widyasari:tosem2024,ferreira:jss2024,kong:saner2022,maddila:esefse2019}\\
	& \cellcolor{lightlightgrey}Does not save time & \cellcolor{lightlightgrey} \cite{strand:icse-sep2020}\\
	& Noise in training data & \cite{li:fse2022,tufano:icse2022,hong:ist2024,chueshev:icsme2020,sghaier:fse2024,jiang:jss2019,asthana:esecfse2019}\\
	& \cellcolor{lightlightgrey}Bias in recommendations & \cellcolor{lightlightgrey} \cite{tecimer:ease2021,mirsaeedi:icse2020,yu:ist2016,strand:icse-sep2020,sharma:spe2019,thongtanunam:saner2015,yu:apsec2014,fan:emse2018,hajari:tse2024,asthana:esecfse2019}\\[0.05cm] \hline

	\multirow{5}{*}{{\bf Evaluation}} & Suboptimal metrics &  \cite{li:fse2022,wang:ist2021b,kartal:infsof2024,chueshev:icsme2020,lu:ase2024,chouchen:emse2023,sulun:icpmdase2019,xia:icsme2015}\\
	& \cellcolor{lightlightgrey}Reliability of oracle & \cellcolor{lightlightgrey} \cite{siow:saner2020,xia:sm2017,qiao:saner2024,rong:icse2022,asthana:esecfse2019,jiang:jcst2015}\\
	& Lack of tradeoffs assessment &  \cite{tao:msr2015,ouni:icsme2016,huq:ist2022,shan:esecfse2022,hajari:tse2024}\\
	& \cellcolor{lightlightgrey}Data leakage & \cellcolor{lightlightgrey} \cite{zhang:ase2022}\\
	& Relevance for practitioners not assessed &  \cite{fregnan:emse2022}\\[0.05cm] \hline
	
	\multirow{4}{*}{{\bf Usability}} & \cellcolor{lightlightgrey}Responsiveness and scalability & \cellcolor{lightlightgrey} \cite{siow:saner2020,fregnan:jss2023,menarini:ase2017,froemmgen:icse2024,guo2020review,wu:kbs2022,ye:ieee2019,maddila:esefse2019}\\
	& Steep learning curve &  \cite{menarini:ase2017,fadhel:iccq2021,yang:emse2024,rahman:icsme2023,froemmgen:icse2024}\\
	& \cellcolor{lightlightgrey}Intepretability of LLMs & \cellcolor{lightlightgrey} \cite{olewicki:fse2024}\\
	& Information overload &  \cite{wang:ase2019,fregnan:jss2023}\\[0.05cm] \hline
	
	\multirow{4}{*}{{\bf Deployment}} & Difficult to integrate in developers' workflow &  \cite{barnett:icse2015}\\
	& \cellcolor{lightlightgrey}Human factors & \cellcolor{lightlightgrey} \cite{egelman:icse2020,rahman:fse2024}\\
	& Privacy concerns &  \cite{hijazi:issre2021}\\
	& \cellcolor{lightlightgrey}Too expensive & \cellcolor{lightlightgrey} \cite{chouchen:emse2024,vijayvergiya:aiware2024,guo:saner2019,lu:apsec2023,nashaat:tse2024}\\[0.05cm]

    \rulec
    \end{tabular}
    }
\end{table}

\tabref{tab:concerns} summarizes the concerns raised and limitations observed by researchers when experimenting with the proposed automated solutions. \tabref{tab:concerns} organizes the identified issues into four parent categories, \emph{performance}-, \emph{evaluation}-, \emph{usability}-, and \emph{deployment}-related issues. For each of them, references to the papers in which we found evidence of such an issue have been reported. In the following, we discuss the main issues identified, focusing on those that are either very popular (\ie reported in several papers) or that provide interesting insights for future work. We use the icon ``\faChevronCircleRight'' to highlight lessons learned and directions for future work.

\subsubsection{Performance-related issues} We use the term ``performance'' to refer to the ability of the technique to provide a proper support for the automation of the targeted task\footnote{This is unrelated to performance aspects such as memory footprint.}. Several researchers highlight the \emph{unsatisfactory recommendations} generated by the experimented techniques, which may make them not ready for developers' adoption. This is a quite crosscutting and expected concern, particularly affecting generative tasks requiring the generation of text or code (\eg \emph{implementing the code change requested by a reviewer}). However, even in classification tasks for which automated support proved quite successful, some researchers raised major concerns about their actual effectiveness: Strand \etal \cite{strand:icse-sep2020} observed that while their approach for reviewer recommendations performed well when evaluated on historical data, it did not seem to save time to developers once deployed in industry. \faChevronCircleRight~ This stresses the importance of experimenting the proposed techniques in realistic scenarios, which can provide feedback about their actual usefulness.

A very popular limitation of the automation techniques is also the \emph{limited support they offer in specific scenarios}. The term ``scenario'' here can have different meanings. For techniques relying on historical data such as those recommending reviewers based on past assignments, or those retrieving code reviews performed in the past for code similar to the one under review, there are limitations related to their applicability on ``previously unseen data''. For example, retrieving reviews from the past does not allow the approach to generate previously unseen review comments \cite{hong:esecfse2022}, something doable nowadays by training DL models. \faChevronCircleRight~ Thus, while retrieval-based techniques offer specific advantages even in the context of generative tasks (\eg they are substantially faster as compared to DL-based techniques), relying on them may be recommendable mostly in quite stable contexts in which the development team, review process, and the code base are not expected to undergo major and continuous changes (thus keeping the value of what learned in the past). One emerging concern is the already mentioned lack of support by DL-based code review automation techniques for low-resource languages \cite{sghaier:fse2024}. It is reasonable to expect that the performance of code review automation techniques substantially drop for these languages, \faChevronCircleRight~ highlighting the importance of investigating the applications of these tools in ``niche usage scenarios'', as recently done in the context of code generation \cite{chen:icpc2022,cassano2023knowledge,van:forge2024}. Also, some researchers highlighted the limited applicability of their technique in specific scenarios which are, however, specific of the tackled problem and experimented technique. For example, Huang \etal \cite{huang:tse2020} claim that their approach to predict the silent class of a commit is unable to deal with tangled commits.

Still related to the performance of the proposed automated techniques, several studies presenting learning-based techniques report the presence of \emph{noise in training data} as a major concern. Given the amount of data on which these techniques rely for the learning, it is difficult to guarantee the quality of training data. For example, when looking at reviewers' recommenders noise can come from developers  using multiple accounts, which are treated as different developers by the approach \cite{jiang:jss2019}, or even from sub-optimal assignment made in the past, \ie the reviewer assigned to a pull request was not the most appropriate, but maybe the one which had a lower workload in that specific moment \cite{asthana:esecfse2019}. Similarly, researchers raised concerns about the quality of training data used for DL models aimed at generating review comments or implementing the code change requested by a reviewer \cite{li:fse2022,tufano:icse2022}. These approaches usually learn from triplets featuring (i) the code submitted for review, (ii) the review comments posted by humans, and (iii) the revised code implementing the changes requested in the review comments. This data is automatically mined from forges such as GitHub and, consequently, can feature noisy data. For example, the collected revised code, while being a modified version of the code submitted for review, may not actually implement the reviewers' comments, but other unrequested changes. The mined triplet will thus ``teach'' something wrong to the DL model. Despite the major cleaning efforts performed by researchers \cite{li:fse2022,tufano:icse2022}, noisy instances survive in the training data, since it is difficult for a single research group to have the man power to manually validate the whole dataset. \faChevronCircleRight~ A joint effort of the research community working on the automation of code review activities would be needed (at least for specific tasks of interest), similarly to what done in other fields like image recognition \cite{quanzeng:aaai16}. 

Finally, researchers working on reviewers' recommendation and predicting code changes approval/merge report the presence of \emph{bias in recommendations} generated by their techniques \cite{tecimer:ease2021,mirsaeedi:icse2020,yu:ist2016,strand:icse-sep2020,sharma:spe2019,thongtanunam:saner2015,yu:apsec2014,fan:emse2018,hajari:tse2024,asthana:esecfse2019}. When it comes to recommend reviewers, bias manifests in the fact that reviewers who have been employed more in the past will also be employed more in the future. The bias becomes even more evident if the approach is re-trained over time to include new data, also featuring pull requests in which the approach has been employed (thus again promoting over and over the same reviewers). When it comes to predicting code changes approval/merge, researchers reported a negative bias of the techniques towards pull requests opened by newcomers. \faChevronCircleRight~ These two examples highlight the importance of considering human factors in the evaluation of the proposed techniques, besides computing performance-related metrics.

\subsubsection{Evaluation-related issues} For this category, we discuss the first three types of concern reported in \tabref{tab:concerns}, since the last two (\ie \emph{data leakage} and \emph{relevance for practitioners not assessed}) have only been reported in one paper each.

The usage of \emph{suboptimal metrics} in the run empirical validations is a major concern for the new line of research tackling generative tasks \cite{li:fse2022,lu:ase2024,kartal:infsof2024}. For example, in the ``generating review comments'' task the technique is provide as input a code to review and it is expected to comment on its quality in natural language as a human would do. The question is how to automatically assess the quality of the generated comments. As explained in the context of RQ$_3$, since the code to review is usually mined from open source projects, researchers usually compute a similarity metric (\eg the BLEU score \cite{papineni:acl2002}) between the generated comment and the comments that were posted by human reviewers for that same code. However, there are many issues with this evaluation procedure. First, two completely different natural language comments may point to the same quality issue in the code. For example, the deep learning model may output ``please rename variable \emph{h} to something more meaningful'' while the human reviewer may write ``change \emph{h} to \emph{height}''. A textual similarity metric between these two comments would point to the low quality of the comment generated by the deep learning model while, in reality, the technique outputted a meaningful recommendation. Second, the model may correctly spot a quality issue which, however, has been missed by the human reviewers, thus not having any ``similar human comment'' to compare with. This again would result in a correct recommendation considered wrong. On top of generative tasks, there are other code review automation tasks for which the metrics used for evaluation only represent a weak proxy for the actual usefulness of the approach. For example, Chueshev \etal \cite{chueshev:icsme2020} stress that evaluation metrics such as top@k accuracy and MRR which they use in the context of reviewer recommendation might not align with the practical use of their technique, since they do not focus on the actual value added by new reviewer recommendations. A strongly-related concern possibly affecting the validity of the reported empirical evaluations is the limited \emph{reliability of oracles}, which mirrors the previously discussed ``\emph{noise in training data}'' but on the ``test data''. \faChevronCircleRight~Future work should aim at (i) defining metrics better capturing the actual usefulness of code review automation techniques, for example assessing the relevance of a generated comment for a given code under review, rather than only comparing it with comments written by humans; and (ii) creating curated benchmarks for code review automation, similarly to what the research community is doing for code generation \cite{hao:icse2024}.

The third evaluation-related issue we discuss concerns to the \emph{lack of tradeoffs assessment}. Concrete examples of this issue may be: (i) focusing the evaluation of a reviewer recommender only on the correctness of the recommendation, without considering the workload distribution among reviewers as one of the objectives to meet \cite{hajari:tse2024}; (ii) not considering that specific decisions may be influenced by interpersonal relationships rather than by objective factors \cite{ouni:icsme2016}; and (iii) ignoring in the evaluation the cost of adopting a novel tool the developers are not familiar with \cite{tao:msr2015}. \faChevronCircleRight~A more comprehensive view of the tradeoffs that come into play when a new code review automation technique is proposed would be desirable in the performed empirical evaluations. However, this may not be doable without running case studies, which are non-trivial to run. At least, a careful discussion of the not-assessed tradeoffs is recommendable for works in the area, especially considering the socio-technical nature of code reviews.

\subsubsection{Usability-related issues} Automation solutions proposed in academia are often implemented in the form of prototypes, with little attention given to non-functional attributes such as \emph{responsiveness and scalability} \cite{siow:saner2020,fregnan:jss2023,menarini:ase2017,froemmgen:icse2024,guo2020review,wu:kbs2022,ye:ieee2019,maddila:esefse2019}. Sometimes these issues are indeed the result of non-optimized code, while in other cases are intrinsic limitations of the proposed approaches. For example, retrieval-based techniques may experience an increasing lack of responsiveness with the growth of the knowledge base from which information is retrieved. Similarly, visualization techniques may not scale to accomodate too complex objects/large amounts of information (\eg a quite large code diff in the context of code review).  

The developed solutions may also be characterized by a \emph{steep learning curve} \cite{menarini:ase2017,fadhel:iccq2021,yang:emse2024,rahman:icsme2023,froemmgen:icse2024}, which is however difficult to assess without human-based studies. Finally, usability concerns may come from \emph{information overload} \cite{wang:ase2019,fregnan:jss2023} and the lack of interpretability of deep learning models \cite{olewicki:fse2024}. Concerning the former, Fregnan \etal \cite{fregnan:jss2023} discuss the risk of information overload when visualizing too intricate merge requests, with the concrete risk of hindering useful information to the reviewers rather than helping them in the code inspection.

\faChevronCircleRight~Since the overall goal of code review automation is to save time to software developers, the usability of the proposed solutions should be considered as a first-class citizen, both at design and evaluation time. Currently, most of techniques are assessed in \emph{in-vitro} evaluations, relying on test sets built by mining software repositories. These evaluations completely neglect the ``usability'' aspect. For some code review tasks for which automation has only been recently targeted (\eg \emph{generating review comments}) this may be reasonable considering that the proposed solutions are still far from generating meaningful recommendations most of times. However, for other tasks such as \emph{recommending reviewers}, dozens of techniques have been proposed with the most recent ones achieving excellent performance on the artificial benchmarks. Investigating their usability becomes now important.

\subsubsection{Deployment-related issues} Related to the former concerns are the deployment-related issues discussed by researchers. We found these issues only discussed in a few papers and mostly pointing to the possibility that the proposed technique may be too expensive to deploy in practice \cite{chouchen:emse2024,vijayvergiya:aiware2024,guo:saner2019,lu:apsec2023,nashaat:tse2024}. This type of concern is mostly related to the proposal of AI-based solutions: Deploying an in-house AI assistant may require substantial monetary investments for training large DL models, making them available on powerful servers, and maintaining them (\eg retraining them) to keep their usefulness over time. \faChevronCircleRight~Given the growth of the AI4SE research field, it is important to step back and also consider the cost-related implications of these techniques, both in terms of money and environmental impact. Integrating techniques such as quantization \cite{gholami2022survey}, knowledge distillation \cite{hsieh2023distilling}, and parameter-efficient fine-tuning \cite{houlsby2019parameter} can help in reducing both the memory footprint and the training/inference cost of the proposed solutions. Still, the impact of these techniques on the performance (quality of recommendations) of code review automation tools must be carefully assessed (\eg a quantized model may experience a substantial lost of performance when compared to the original model). 

%!TEX root = main.tex

%%%%%%%%%%%%%%%%%%%%%%
\section{Threats to Validity} 
\label{sec:threats}
%%%%%%%%%%%%%%%%%%%%%%

\textbf{Threats to construct validity} concern the relation between theory and observation. We only included papers indexed in the six queried databases. Also, we only focused on works published in software engineering venues. Thus, there might be additional studies we missed. The snowballing procedure we applied helps in mitigating this threat, despite the fact that we only performed one round of snowballing. We believe that most relevant studies were included based on the expertise the authors have in this domain. Also, the number of papers included in our SLR is large enough to answer our research questions and the main findings are unlikely to change even assuming a few missing works. Also, as a design decision, we did not apply any quality assessment criteria to exclude studies from our SLR. Indeed, we felt that the subjectiveness of this judgement was too high and decided to consider peer-reviewed papers as a sort of ``automated quality filter''. We acknowledge that some peer-reviewed studies included in our study might feature flaws or wrong claims which could also potentially affect our findings (\eg wrong data extracted).

\textbf{Threats to internal validity} concern external factors we did not consider that could affect the variables being investigated. The search engines we used are continuously updated, both in terms of search features as well as in terms of papers they index. We cannot ensure replicability of our findings. However, we provide all material we collected in an online appendix \cite{replication}.

\textbf{Threats to external validity} concern the generalizability of our findings.  We decided to focus our SLR only on the literature proposing code review automation tools ignoring, for example, the body of knowledge related to empirical studies on code review. Furthermore, as our paper search was conducted in December 2024, our SLR misses works which have been later indexed in the searched database.

\textbf{Threats to conclusion validity} concern the relations between the conclusions and our analyzed data. The main threat here is related to the correctness of the data we extracted from the inspected papers. To minimize errors, the two authors always double-checked the information each of them collected. However, especially in the context of RQ$_5$, we felt that a strong subjectivity component was involved in deciding what should be considered as a limitation/concern discussed by the paper's authors. We acknowledge that we may have missed several insights reported in the read works. Still, we feel the set of limitations/concerns discussed in RQ$_5$ to be quite representative of those we encountered while reading papers for this SLR.
%!TEX root = main.tex

%%%%%%%%%%%%%%%%%%%%%%
\section{Conclusions} 
\label{sec:conclusion}
%%%%%%%%%%%%%%%%%%%%%%
We presented a systematic literature review involving \selected papers presenting solutions for the automation of code review-related tasks. Firstly, we categorized the \numberOfTasks tasks for which at least one automated approach has been proposed. We then summarized the under-the-hood solutions behind these approaches, and the metrics used in their empirical evaluation. We also looked for the presence of replication packages in the \selected papers, checking for the available ones whether they are still reachable and if they provide access to the presented approach and the used data. In the end, we highlighted the concerns and limitations researchers discussed when presenting and evaluating the proposed approaches, using them to highlight possible directions for future work.

We release the raw data summarized in the SLR in our online appendix \cite{replication}.

%%%%%%%%%%%%%%%%%%%%%%%%%%%%%%%%%%%%%%%%
% Acknowledgments, disabled for now. 
%%%%%%%%%%%%%%%%%%%%%%%%%%%%%%%%%%%%%%%
\begin{acks}
This project has received funding from the European Research Council (ERC) under the European Union's Horizon 2020 research and innovation programme (grant agreement No. 851720).
\end{acks}

\bibliographystyleP{ACM-Reference-Format}
\bibliographystyle{unsrt}
\bibliography{reference}

%!TEX root = main.tex

%%%%%%%%%%%%%%%%%%%%%%
%%%%%%%%%%%%%%%%%%%%%%
\appendix
\section{Appendix}  \label{sec:appendix}
%%%%%%%%%%%%%%%%%%%%%%
%%%%%%%%%%%%%%%%%%%%%%

\begin{table}[h!]
	\centering
    \caption{Venue names}
    \label{tab:venue_names}
    {\footnotesize
    \begin{tabular}{cl}
    \toprule
    {\bf Acronym} & {\bf Venue Name}\\\midrule
    ASE               & International Conference on Automated Software Engineering\\
    COMPSAC    & Annual Computer Software and Applications Conference\\
    EASE            & International Conference on Evaluation and Assessment in Software Engineering\\
    EMSE           & Empirical Software Engineering\\
    ESEC/FSE    & European Software Engineering Conference and Symposium on the Foundations of Software Engineering\\
    ESEM           & International Symposium on Empirical Software Engineering and Measurement\\
    ICSE             & International Conference on Software Engineering\\
    ICSE-SEIP    & International Conference on Software Engineering: Software Engineering in Practice\\
    ICSME          & International Conference on Software Maintenance and Evolution\\
    IEEE Access & IEEE Access\\
    ISSRE           & International Symposium on Software Reliability Engineering\\
    IST                & Journal of Information and Software Technology  \\
    JSS               & Journal of Systems and Software\\
    MSR             & International Conference on Mining Software Repositories\\
    PACMSE      & Proceedings of the ACM on Software Engineering\\
    PROMISE    & International Conference on Predictive Models and Data Analytics in Software Engineering\\
    SANER        & International Conference on Software Analysis, Evolution and Reengineering\\
    SEKE          &  International Conference on Software Engineering and Knowledge Engineering\\
    TOSEM       & Transactions on Software Engineering and Methodology\\
    TSE             & Transactions on Software Engineering\\
    \bottomrule
    \end{tabular}
    }
\end{table}

%%
%% If your work has an appendix, this is the place to put it.

\end{document}